\documentclass[twocolumn, tighten, times, twocolappendix]{aastex631}
\let\tablenum\relax
\usepackage{siunitx}
\usepackage{amsmath}
\usepackage{mathtools}
\usepackage{microtype}

\graphicspath{{figures/}}

\DeclarePairedDelimiter\p{(}{)}

\newcommand{\lp}{\left(}
\newcommand{\rp}{\right)}
\newcommand{\ls}{\left[}
\newcommand{\rs}{\right]}
\newcommand{\labs}{\left|}
\newcommand{\rabs}{\right|}
\newcommand{\lb}{\left\{}
\newcommand{\rb}{\right\}}

\newcommand{\mjup}{\ensuremath{M_\text{Jup}}}

\DeclareSIUnit{\year}{yr}
\DeclareSIUnit{\day}{d}
\DeclareSIUnit{\dex}{dex}
\DeclareSIUnit{\gauss}{G}

\ifx\qty\undefined
\newcommand{\qty}{\SI}
\else
\fi

\ifx\unit\undefined
\newcommand{\unit}{\si}
\else
\fi

\shorttitle{Hydrodynamics and survivability during post-main-sequence planetary engulfment}
\shortauthors{Yarza et al.}

\begin{document}

\title{Hydrodynamics and survivability during post-main-sequence planetary engulfment}

\author[0000-0003-0381-1039]{Ricardo Yarza}
\altaffiliation{NASA FINESST Fellow}
\altaffiliation{Frontera Computational Science Fellow}
\affiliation{Department of Astronomy and Astrophysics, University of California, Santa Cruz, CA 95064, USA}
\affiliation{Texas Advanced Computing Center, University of Texas, Austin, TX 78759, USA}

\author{Naela B. Razo-L\'{o}pez}
\affiliation{Department of Astronomy and Astrophysics, University of California, Santa Cruz, CA 95064, USA}

\author[0000-0003-2333-6116]{Ariadna~Murguia-Berthier}
\altaffiliation{NASA Hubble Fellow}
\affiliation{Department of Astronomy and Astrophysics, University of California, Santa Cruz, CA 95064, USA}
\affiliation{Center for Interdisciplinary Exploration and Research in Astrophysics (CIERA), 1800 Sherman Ave., Evanston, IL 60201, USA}

\author[0000-0001-5256-3620]{Rosa Wallace Everson}
\altaffiliation{NSF Graduate Research Fellow}
\affiliation{Department of Astronomy and Astrophysics, University of California, Santa Cruz, CA 95064, USA}

\author[0000-0003-3062-4773]{Andrea Antoni}
\altaffiliation{NSF Graduate Research Fellow}
\affiliation{Department of Astronomy, University of California, Berkeley, CA 94720, USA}

\author[0000-0002-1417-8024]{Morgan MacLeod}
\affiliation{Center for Astrophysics, Harvard \& Smithsonian, 60 Garden Street, Cambridge, MA 02138, USA}

\author[0000-0001-7493-7419]{Melinda Soares-Furtado}
\altaffiliation{NASA Hubble Fellow}
\affiliation{Department of Astronomy, University of Wisconsin-Madison, 475 N. Charter St., Madison, WI 53703, USA}

\author[0000-0002-8229-3040]{Dongwook Lee}
\affiliation{Department of Applied Mathematics and Statistics, University of California, Santa Cruz, CA 95064, USA}

\author[0000-0003-2558-3102]{Enrico~Ramirez-Ruiz}
\affiliation{Department of Astronomy and Astrophysics, University of California, Santa Cruz, CA 95064, USA}

\correspondingauthor{Ricardo Yarza}
\email{ryarza@ucsc.edu}

\begin{abstract}
    The engulfment of substellar bodies (SBs, such as brown dwarfs and planets) by giant stars is a possible explanation for rapidly rotating giants, lithium-rich giants, and the presence of SBs in close orbits around subdwarfs and white dwarfs. We simulate the flow in the vicinity of an engulfed SB in three-dimensional hydrodynamics. We model the SB as a rigid body with a reflective surface because it cannot accrete. This reflective boundary changes the flow morphology to resemble that of engulfed compact objects with outflows. We measure the drag coefficients for the ram pressure and gravitational drag forces acting on the SB, and use them to integrate its trajectory inside the star. We find that engulfment can increase the luminosity of a \(1M_\odot\) star by up to a few orders of magnitude. The time for the star to return to its original luminosity is up to a few thousand years when the star has evolved to \(\approx10R_\odot\) and up to a few decades at the tip of the red giant branch. No SBs can eject the envelope of a \(1M_\odot\) star before it evolves to \(\approx10R_\odot\), if the orbit of the SB is the only energy source contributing to the ejection. In contrast, SBs as small as \(\approx10M_\text{Jup}\) can eject the envelope at the tip of the red giant branch. The numerical framework we introduce here can be used to study planetary engulfment in a simplified setting that captures the physics of the flow at the scale of the SB\@.
\end{abstract}

\section{Introduction}
Common-envelope evolution \citep[hereafter CEE;][]{Paczynski1976} is a process in which a star engulfs a companion (substellar or otherwise). The known planetary system architectures imply that a large fraction of planets and brown dwarfs (hereafter substellar bodies, SBs) will eventually undergo CEE \citep{Villaver2009,Mustill2012,Nordhaus2013,Schlaufman2013,Sun2018}.
Throughout this work, we will refer to CEE between a star and an SB as ``planetary engulfment,'' and use ``CEE'' for the more general interaction between a star and a companion of any mass.

Planetary engulfment is a possible explanation for several unsolved problems in stellar and planetary system evolution. Observations have found SBs in close orbits around subdwarfs and white dwarfs \citep[][for a summary see \citealt{Kruckow2021}]{Schmidt2005,Littlefair2006,Maxted2006,Littlefair2007,Silvestri2007,Littlefair2008,Geier2009,Charpinet2011,Breedt2012,Casewell2012,Liu2012,RebassaMansergas2012,Beuermann2013,Steele2013,McAllister2015,Almeida2017,Schaffenroth2015,Parsons2017,Pala2018,Casewell2020,Vanderburg2020,Schaffenroth2021,vanRoestel2021}. These systems might have reached their current orbital configurations dynamically through the Kozai--Lidov mechanism \citep{Kozai1962,Fabrycky2007,Katz2011,Naoz2012,Socrates2012,Shappee2013,Munoz2020,OConnor2021} or via an engulfment phase in which the SB ejected the envelope of the star that engulfed it \citep{Livio1984,Nelemans1998,Lagos2021,Merlov2021,Zorotovic2022}. During engulfment, orbital energy dissipation shrinks the orbit of the system significantly.
Even if the SB does not survive, engulfment might result in an isolated white dwarf with \( \gtrsim\unit{\mega\gauss} \) magnetic fields \citep{Nordhaus2011,Guidarelli2019}.

Engulfment could explain observations of anomalous rotation among some giant stars. During engulfment, the SB transfers the angular momentum of its orbit into the stellar envelope, resulting in either enhanced or reduced rotation, depending on the alignment of the angular momentum vectors of the star and the orbit. Engulfment can speed up the surface of giant stars up to the observed values, and even up to a significant fraction of their critical speeds \citep{Peterson1983,Soker1998,Siess1999,Zhang2014,Privitera2016,Privitera2016a,Qureshi2018,Stephan2020}.

While the stellar surface abundance of the \( ^7 \)Li isotope generally decreases throughout stellar evolution \citep{Bodenheimer1965,Deliyannis2000,Piau2002,Baumann2010,Monroe2013,Melendez2014,Carlos2016,Carlos2019,SoaresFurtado2021}, dropping significantly at the onset of the first dredge-up phase, \( \approx1\% \) of giants have abundances \( \geq\qty{1.5}{\dex} \) \citep[e.g.,][]{Wallerstein1982,Brown1989,Balachandran2000,Charbonnel2000,Reddy2005,Carlberg2010,Charbonnel2010,Kumar2011,Martell2013,Adamow2014,Adamow2015,Yan2018,Li2018,Deepak2019,Gao2019,Singh2019}. Moreover, \( \approx6\% \) of these \( ^7 \)Li-rich giants exceed the meteoritic abundance of \qty{3.3}{\dex}, indicating that additional \( ^7 \)Li must have been generated or deposited within them \citep{Balachandran2000,Zhou2019,Singh2019}. The engulfment of SBs is a possible explanation for high surface \( ^7 \)Li abundances \citep{Sandquist1998,Siess1999,Sandquist2002,AguileraGomez2016,AguileraGomez2016a, SoaresFurtado2021} because SBs do not reach the requisite temperatures to burn their primordial \( ^7 \)Li. However, there are other pathways for lithium enrichment, such as the \cite{Cameron1971} mechanism, which acts after the early red giant branch (RGB). The existence of these different pathways makes it harder to identify the source of enrichment for stars after the early RGB\@. Infrared excess is a potential indicator of stellar mass loss from engulfment, and evolved stars with infrared excess tend to have increased \( ^7 \)Li and rotation rates \citep{Mallick2022}.

Several analytical studies \citep{Metzger2012,Yamazaki2017,Jia2018} have focused on planetary engulfment by main-sequence (MS) or pre-main-sequence stars, where envelope ejection is unlikely because of the high gravitational binding energy. As for post-MS planetary engulfment, early analytical estimates \citep{Nelemans1998,Livio1984} suggest that SBs with masses\footnote{We use the International Astronomical Union nominal values for solar system constants \citep{Prsa2016}.} \( \lesssim10M_\text{Jup} \) cannot unbind the stellar envelope. \citet{Staff2016} simulated the engulfment of a massive planet by stars in the RGB and AGB using 3D hydrodynamics. However, their results regarding envelope ejection were limited by numerical resolution. Overall, planetary engulfment remains a relatively unexplored problem in the context of hydrodynamical simulations.

Previous work on CEE has used the ``wind tunnel'' numerical formalism to study the flow in the vicinity of the engulfed companion, accounting for the density gradient in the stellar envelope \citep[][]{MacLeod2015,MacLeod2015a,MacLeod2017,MurguiaBerthier2017,De2020,Everson2020}. These density gradients change the flow morphology and give angular momentum to the gravitationally focused gas, thereby changing the drag forces on the companion. Most of this previous work has focused on interactions between an evolved star and a compact companion, for which gravitational drag dominates. For substellar companions, ram pressure drag might dominate, depending on stellar structure and on the companion. While some studies have recognized the importance of ram pressure \citep{Staff2016,Jia2018}, it has not yet been accounted for in detail.

\begin{figure}[t]
    \centering
    \includegraphics[width=\columnwidth]{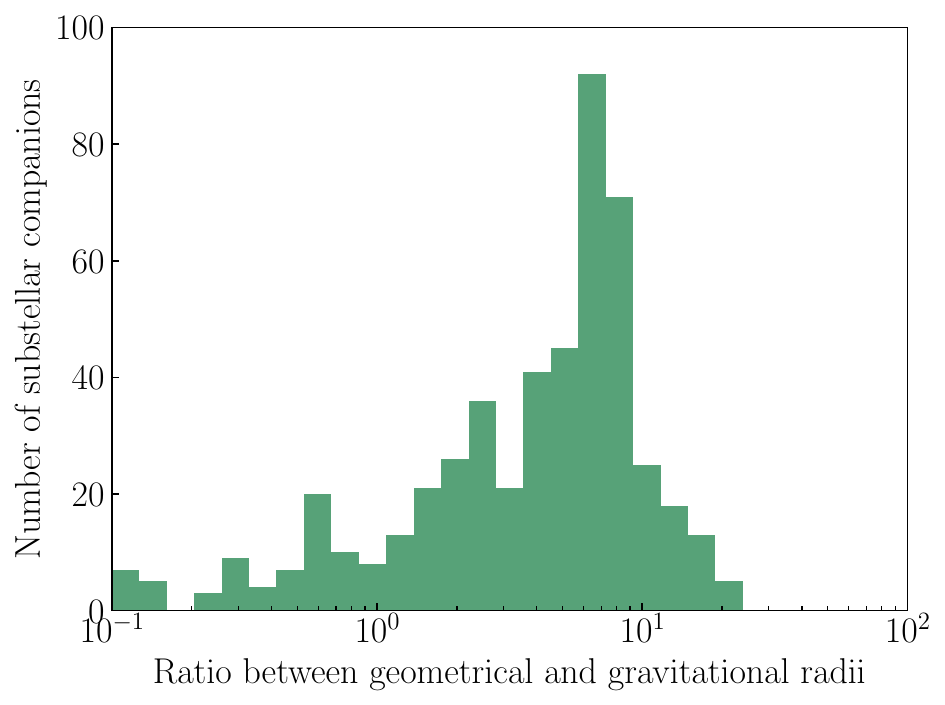}
    \caption{Distribution of exoplanets more massive than Jupiter \citep{exoplanetarchive} over the ratio of their geometrical and gravitational radii at the onset of engulfment, assuming they are engulfed at their current separations. Since tidal decay will lead to the engulfment of planets at smaller orbital separations than their current ones, this value is likely a lower limit. The geometrical radii are larger for \( 82\% \) of these exoplanets.}\label{fig:rpra}
    \end{figure}

\begin{figure*}[t!]
    \centering
    \includegraphics[width=0.495\textwidth]{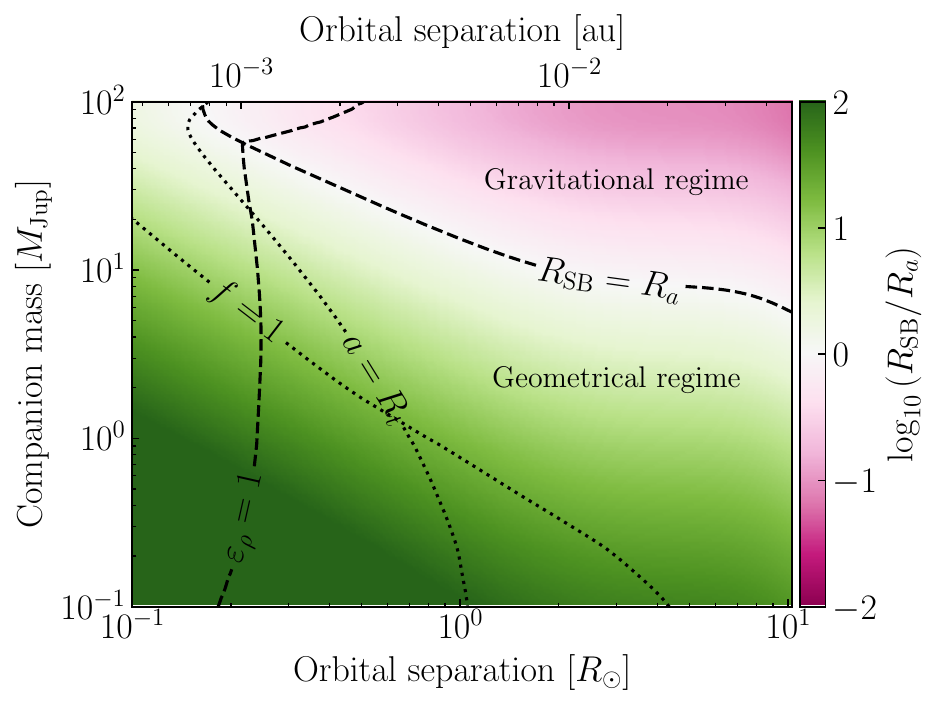}
    \includegraphics[width=0.495\textwidth]{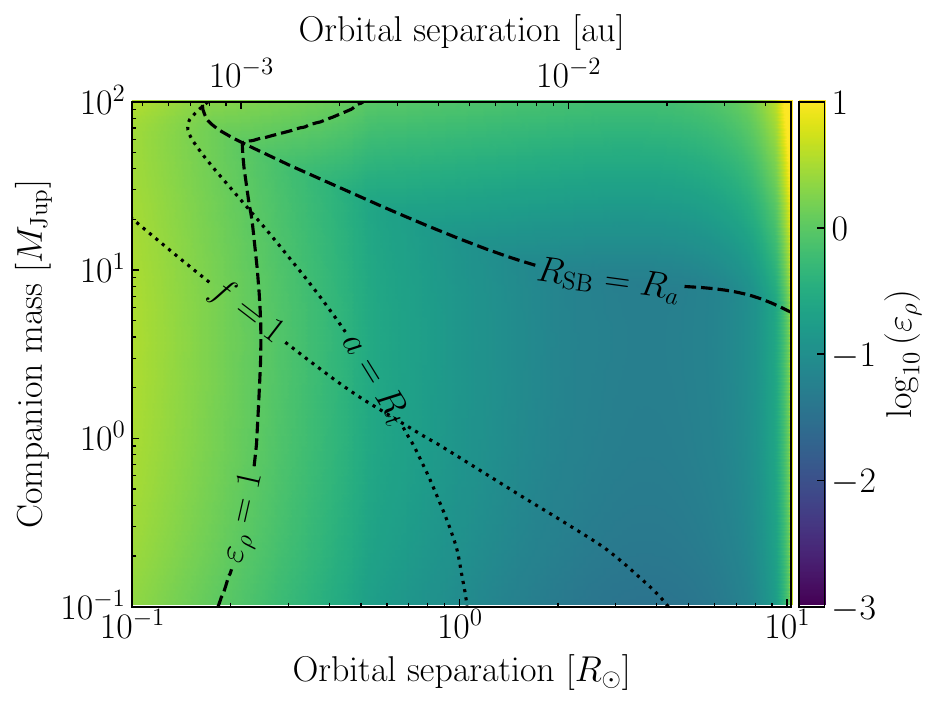}
    \includegraphics[width=0.495\textwidth]{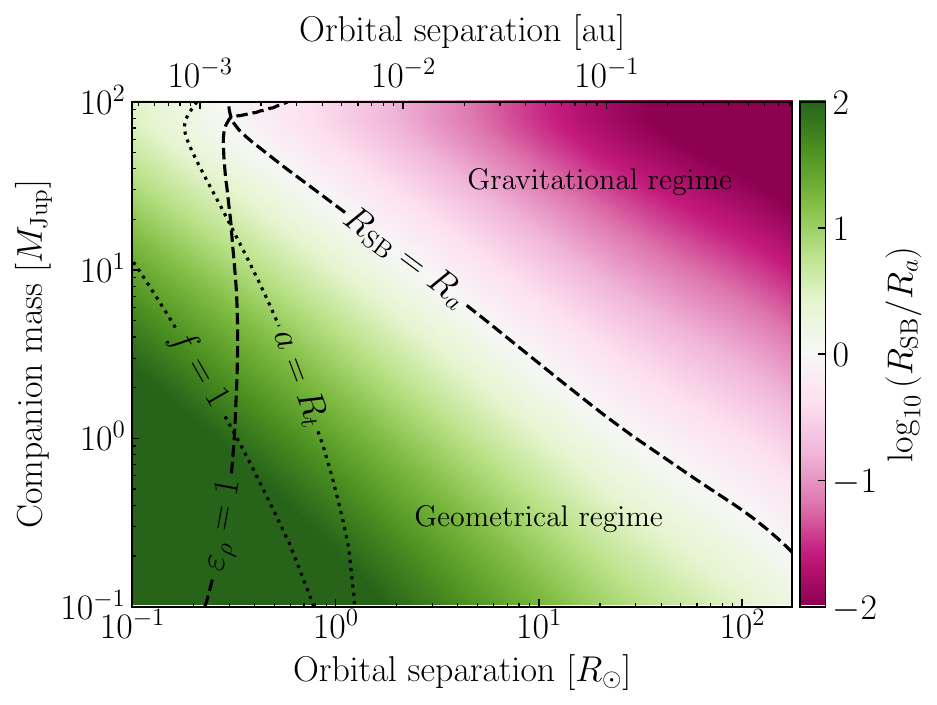}
    \includegraphics[width=0.495\textwidth]{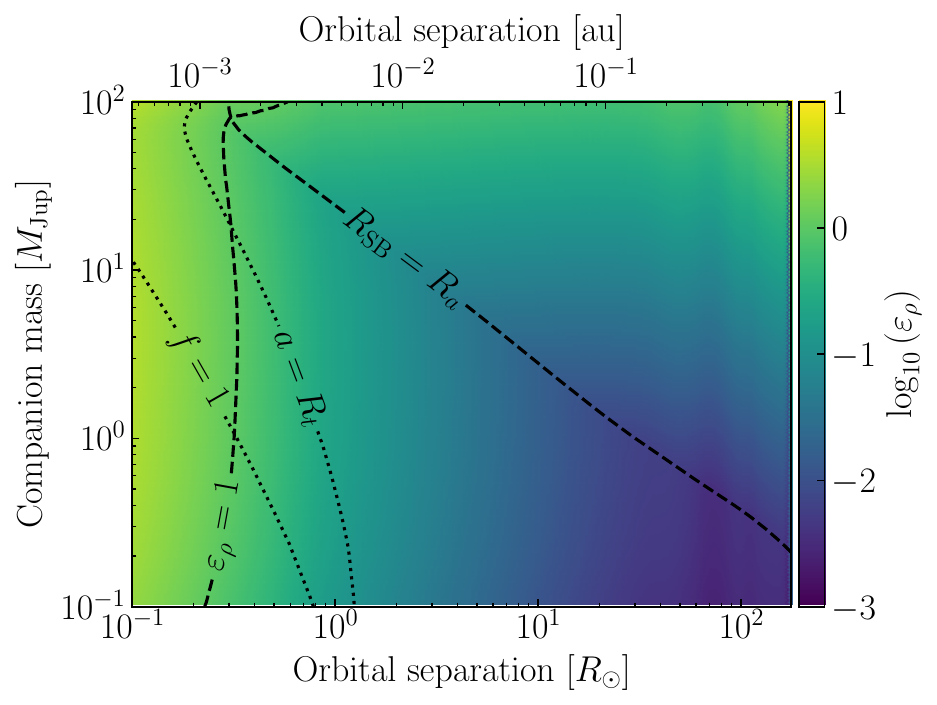}\\
    \caption{Dimensionless flow parameters (left: ratio between geometrical and gravitational radii, right: number of density scale heights across the larger radii) as a function of substellar body (SB) mass and position inside a model of the Sun evolved to \( 10R_\odot \) (top) and to the tip of the red giant branch (bottom). Dashed lines show the transition between geometrical and gravitational regimes, and between mild and strong density gradients. Dotted lines show estimates for SB disruption, either by tidal forces (when the separation equals the tidal radius \( R_t \)) or by ram pressure (\(f=1 \), equation~\ref{eq:disruption:ram}).}\label{fig:flow_params}
    \end{figure*}

The drag forces acting on the engulfed SB influence the dynamics of the orbital decay, the observational signatures associated with it, and ultimately whether the SB can survive engulfment. Here we study planetary engulfment using the wind tunnel numerical formalism.%

In Section~\ref{sec:landscape}, we discuss the physical processes relevant to engulfment, particularly the relative importance of ram pressure and gravitational drag forces. In Section~\ref{sec:cewt}, we provide a brief review of the wind tunnel framework, and discuss its applicability and limitations in the context of planetary engulfment. We discuss flow morphology and drag coefficients in Section~\ref{sec:morphology}, and apply these results to planetary engulfment in Section~\ref{sec:applications}. Section~\ref{sec:conclusion} summarizes our results.

\section{Physical aspects of engulfment}\label{sec:landscape}
\subsection{Gravitational and geometrical regimes}
The interactions of an engulfed SB with the surrounding stellar material are gravitational (gravitational drag) and geometrical (ram pressure drag on the geometrical surface of the SB). Gravitational drag arises from gravitational focusing of material behind the SB; this focused material exerts a force against the direction of motion. An engulfed SB with mass \( M_\text{SB} \) travels relative to the surrounding gas at an orbital speed \( v_\text{orb} \). Gas with an impact parameter smaller than the gravitational radius of the SB,
\begin{equation}
\label{eq:ra}
R_a=2GM_\text{SB}/v_\text{orb}^2,
\end{equation}
will be gravitationally focused behind the SB\@. This gas will exert a force \( F_g = \pi C_g R_a^2 \rho v_\text{orb} ^2 \) \citep{Hoyle1939,Bondi1944,Bondi1952}, where \( G \) is the gravitational constant \citep[we use the value given in][]{Tiesinga2021}, \( \rho \) is the envelope mass density, and \(C_g\) is a dimensionless coefficient of order unity. On the other hand, the pressure field at the surface of the SB will exert a ram pressure force of the form \(F_p=\pi C_p R_\text{SB}^2 \rho v_\text{orb} ^2\), where \(R_\text{SB}\) is the geometrical radius of the SB, and $C_p$ is a dimensionless coefficient of order unity.

Drag forces are approximately equal to the momentum per unit time passing through the cross-section for the corresponding interaction (geometrical or gravitational).
The ratio between ram pressure and gravitational forces is therefore equal to the ratio of the cross-sections, \( \lp R_\text{SB} / R_a \rp ^ 2 \), or equivalently the ratio \( \lp v_\text{orb}/v_\text{esc,SB}\rp ^2 \), where \( v_\text{esc,SB}=\sqrt{2G M_\text{SB}/R_\text{SB}} \) is the escape velocity from the SB\@.

CEE studies have dealt almost exclusively with the engulfment of a compact object, such as a neutron star or black hole, in which case the interactions between the companion and the stellar material are mostly gravitational. Figure~\ref{fig:rpra} shows the ratio between geometrical and gravitational cross-sections at the onset of engulfment for the known exoplanets, assuming they are engulfed at their current orbital separations. Planets are likely to be engulfed at separations smaller than their current separations as a result of tidal decay. Since the Keplerian speed is greater at smaller separations, the gravitational radii of planets is likely to be smaller at engulfment than it is today, and more planets will be in the geometrical regime at the onset of engulfment. Equivalently, planets engulfed at earlier stages of stellar evolution are more likely to be in the geometrical regime, because they must orbit their host star more closely to be engulfed during earlier stages.

Figure~\ref{fig:flow_params} shows the same ratio as Figure~\ref{fig:rpra}, but as a function of SB mass and position inside a \( \mathrm{M_\odot} \) star evolved to \( 10R_\odot \) (top panel) and at the tip of the RGB (bottom panel). We computed the properties of this star using the Modules for Experiments in Stellar Astrophysics \citep[MESA,][]{Paxton2011}. As the SB dives deeper into the envelope, its interactions with the gas become increasingly geometrical because the Keplerian speed increases inwards. While it is possible for the Keplerian speed to decrease inwards if the enclosed mass profile is sufficiently steep, the post-MS envelopes we consider here are extended enough that the Keplerian speed always increases inwards. Therefore, interactions between the SB and the surrounding material become increasingly geometrical throughout engulfment.

Figure~\ref{fig:flow_params} also shows that more massive SBs tend to be deeper in the gravitational regime. Between \( 10^{-1}M_\text{Jup}\) and \( 10^2M_\text{Jup} \), SB radius varies only between \( 0.7R_\text{Jup} \) and \( 1.2R_\text{Jup}\) (we determine the radius of each SB using the mass-radius relations from \citealt{Fortney2007} and \citealt{Chabrier2009}). As \(M_\text{SB}\) increases, $R_\text{SB}$ remains approximately constant, but $R_a$ is proportional to $M_\text{SB}$ (equation \ref{eq:ra}), so \( R_\text{SB}/R_a \) is approximately inversely proportional to $M_\text{SB}$.

The right panels in Figure~\ref{fig:flow_params} shows the number of density scale heights across the SB,
\begin{equation}\label{eq:eps_rho}
    \varepsilon_\rho\equiv\max\lp R_\text{SB},R_a\rp/H_\rho,
\end{equation}
where the density scale height \( H_\rho\equiv-\rho/\lp d\rho/dr\rp \). The dimensionless density gradient \( \varepsilon_\rho \) quantifies the heterogeneity of the flow at the scale of the SB\@. Engulfed SBs typically experience mild (\( 0\leq\varepsilon_\rho\lesssim1 \)) density gradients.

\begin{figure}[t!]
    \centering
    \includegraphics[width=\columnwidth]{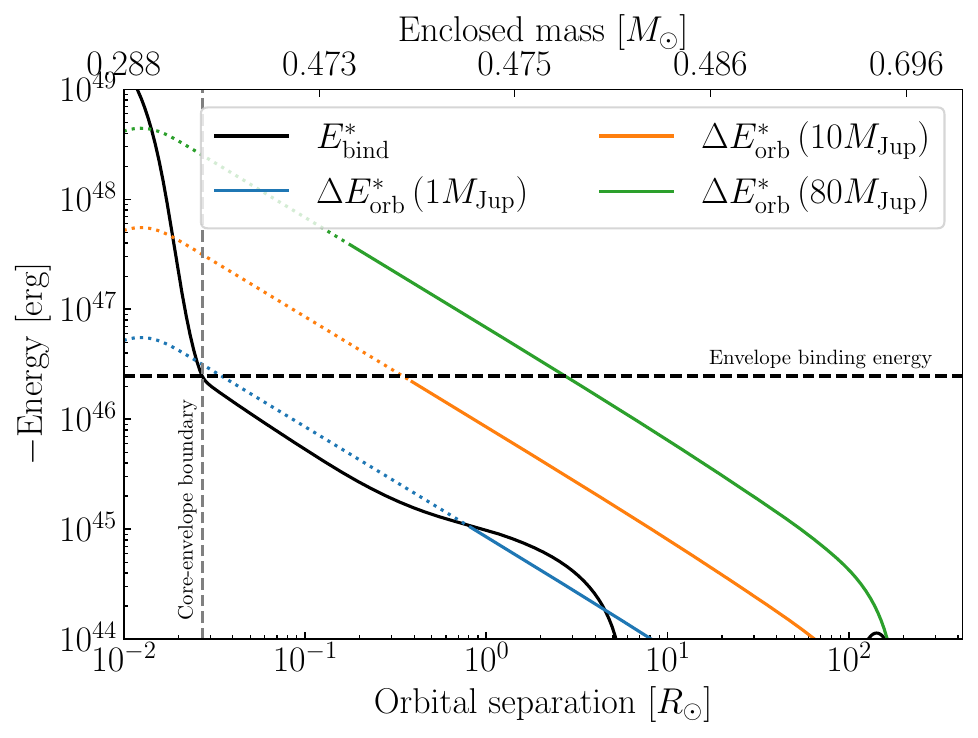}
    \caption{Comparison between the binding energy of exterior material \( E_\text{bind}^* \) and orbital energy \( \Delta E_\text{orb}^* \) deposited by substellar bodies (SBs) of different masses, as a function of position inside a model of the sun at the tip of the red giant branch. Orbital energy deposition lines turn dashed at the estimated destruction point for each SB\@.}\label{fig:alpha}
    \end{figure}

\subsection{Destruction of the substellar body}\label{sec:landscape:destruction}
The SB will be tidally disrupted when
\begin{equation}
    \label{eq:disruption:tides}
    \langle\rho_\text{enc}\rangle = \langle\rho_\text{SB}\rangle,
\end{equation}
where \( \langle\rho_\text{enc}\rangle \) is the average density of the material enclosed by the SB's orbit, and \( \langle\rho_\text{SB}\rangle \) is the average density of the SB\@. This criterion is equivalent to the orbital separation being equal to the tidal radius of the SB.

We estimate the SB will be disrupted by ram pressure when the kinetic energy per unit volume of the incoming flow equals the binding energy per unit volume of the SB, i.e., when \citep{Jia2018}
\begin{equation}
    f \equiv \frac{\rho v_\text{orb}^2}{\langle\rho_\text{SB}\rangle v_\text{esc,SB}^2}=1.
    \label{eq:disruption:ram}
\end{equation}
As shown in Figure~\ref{fig:flow_params}, tidal disruption is the dominant destruction process for most SBs. Ram pressure will disrupt SBs with masses \( \lesssim1M_\text{Jup} \) engulfed early in the red giant branch.    

\subsection{Envelope ejection}
We use the standard energy formalism of CEE \citep{vanDenHeuvel1976,Webbink1984} to estimate analytically whether envelope ejection is possible. In this formalism, an engulfed companion will eject material exterior to the orbital separation \( r \) if the binding energy of that material is smaller in magnitude than the change in orbital energy of the SB, i.e., \( \left|\Delta E_\text{orb}^{*}\right| > \left|E_\text{bind}^*\right| \), where
\begin{gather}
\Delta E_\text{orb}^{*} = 
- \frac{G M_\text{enc} M_\text{SB}}{2r} + \frac{G M_{\text{enc},0} M_\text{SB}}{2r_0},\\
E_\text{bind}^* = \int_{M_\text{enc}}^{M_{\text{enc},0}} \lp - \frac{G M'_\text{enc}}{r} + u \rp dM'_\text{enc}.
\end{gather}
Here, \( u \) is the specific internal energy, \( M_\text{enc} \) is the enclosed mass at orbital separation \( r \), and the \( 0 \) subscript refers to values at the initial time \( t=0 \). These two equations neglect the binding energy between the star and the SB\@; at the low mass ratios we study here, the envelope is much more bound to the core than to the SB, so we omitted these terms for conciseness and readability. Figure~\ref{fig:alpha} compares \( E_\text{bind}^* \) for a MESA model of the Sun at the tip of the RGB to \( \Delta E_\text{orb}^* \) for SBs of several masses. We define the core-envelope boundary as the radial coordinate at which the hydrogen mass fraction is \( 1/10 \). We consider envelope ejection to be possible if the \( \left|\Delta E_\text{orb}^{*}\right| \) is at any point (before destruction of the SB) above the value of the \( \left|E_\text{bind}^*\right| \) curve at the core-envelope boundary. The Figure shows this value as a horizontal line labeled ``envelope binding energy.'' This prescription assumes that energy can be efficiently distributed in the envelope, so that energy deposited at \( r \) can eject material at \( r' < r \). As we will see (Figure~\ref{fig:wind_tunnel_grid}), the SB significantly disturbs material within a few \( \max\lp R_\text{SB}, R_a \rp \) of its current location. Tidal forces destroy massive SBs at orbital separations comparable to their size (see the lines for \( 10\mjup \) and \( 80\mjup \) in Figure~\ref{fig:alpha}), so we expect at least the energy they deposit near the destruction point to reach the core-envelope boundary.
\begin{figure}[t!]
    \centering
    \includegraphics[width=\columnwidth]{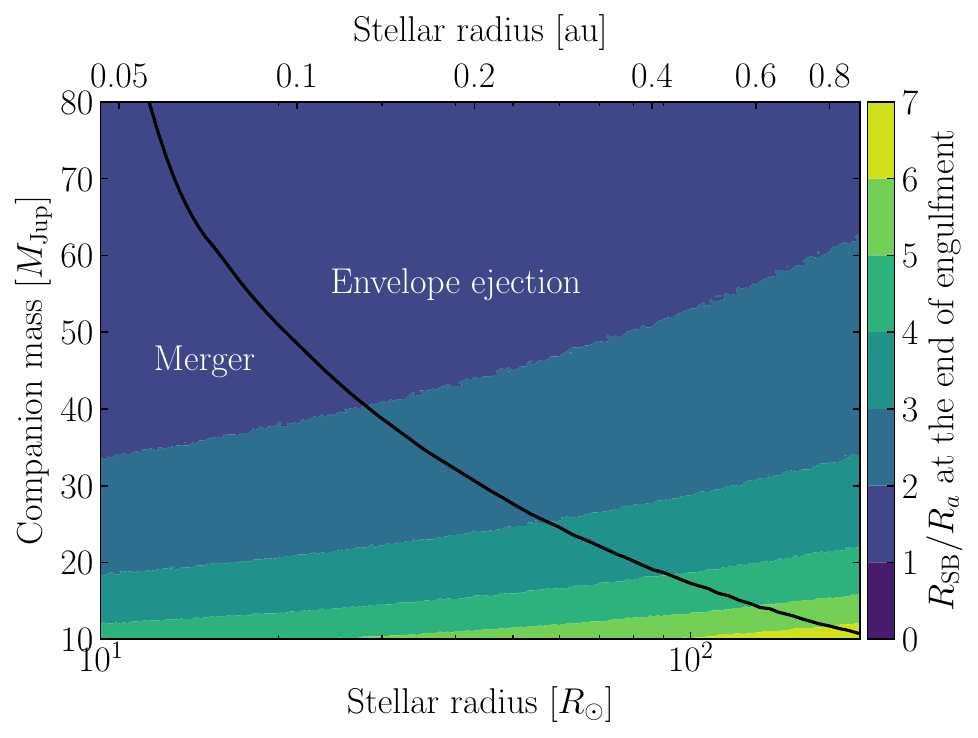}
    \caption{Ratio of geometrical and gravitational cross-sections at the point of envelope ejection or SB destruction, as a function of stellar radius and SB mass. The dashed line shows the minimum mass required to eject the envelope according to the standard energy formalism. When engulfment ends, either by destruction of the SB or envelope ejection, all SBs are in the geometrical regime.}\label{fig:eject}
    \end{figure}

According to the standard energy formalism, objects with masses \( \gtrsim10M_\text{Jup} \) might eject the envelope of a Sun-like star at the tip of the red giant branch. Figure~\ref{fig:eject} shows \( R_\text{SB}/R_a \) at the location of either SB destruction or envelope ejection, as a function of SB mass and stellar radius. The solid line shows the minimum SB mass required to eject the envelope. The star expands throughout the RGB, so its binding energy decreases and envelope ejection by smaller SBs becomes possible. No companion with mass \( <80M_\text{Jup} \) can eject the envelope of an \( M_\odot \) star early in the RGB\@. All SBs are in the geometrical regime when they are destroyed or when they eject the envelope, highlighting the need for numerical models of ram pressure drag.

\subsection{Orbital decay timescales}
The rate of orbital energy dissipation is the work per unit time done by the drag forces, so the orbital decay timescale for ram pressure drag is
\begin{equation}
    \label{eq:t_decay_ram}
    \tau_\text{insp,p} \approx\frac{E_\text{orb}}{\dot{E}_\text{orb}}= \frac{G M_\text{SB}M_\text{enc}}{2a F_p v_\text{orb}}=\frac{G M_\text{enc} M_\text{SB}}{2\pi C_p a R_\text{SB}^2 v_\text{orb}^3 \rho},
\end{equation}
and for gravitational drag
\begin{equation}
    \label{eq:t_decay_grav}
    \tau_\text{insp,g} \approx \frac{G M_\text{SB}M_\text{enc}}{2a F_g v_\text{orb}}=\lp\frac{M_\text{enc}}{M_\text{SB}}\rp\frac{v_\text{orb}}{8\pi G C_g a \rho}.
\end{equation}
In the geometrical regime, the orbits of more massive SBs decay more slowly because they experience approximately the same force, but have larger inertia. As before, radius is almost constant in mass between \( 1M_\text{Jup} \) and \( 100M_\text{Jup} \), so the change in the geometrical orbital decay timescale as a result of changes in SB radius is negligible. In the gravitational regime, however, the orbits of more massive SBs decay faster because the gravitational cross-section scales as the square of their mass, overcoming the inertial term. Whether the orbital decay timescale is equation~\ref{eq:t_decay_ram} or equation~\ref{eq:t_decay_grav} depends on \( R_\text{SB}/R_a \) at the onset of engulfment.

Tides might dominate the orbital decay of the companion in the outer envelope \citep{Stephan2020}, but we do not account for them here. Additionally, the drag orbital decay timescales are sensitive to the density of the stellar envelope at the onset of engulfment because the object will spend most of its time in the outer envelope, where drag forces are smaller. The orbital and tidal evolution during the post-main-sequence set the initial conditions for engulfment. Hydrodynamical simulations \citep{Staff2016} show that the star will overflow its Roche lobe. Our calculations neglect these effects by using the unperturbed stellar structure throughout engulfment.

We expect the orbits of all SBs engulfed near the tip of the RGB to decay on a timescale comparable to Equation~\ref{eq:t_decay_grav} because they are in the gravitational regime at the onset of engulfment (see Figure~\ref{fig:flow_params}). In earlier stages of stellar evolution, however, less massive SBs can be in the geometrical regime at the onset of engulfment, and more massive SBs in the gravitational regime. Since the gravitational and ram pressure timescales have opposite scaling in mass, the orbits of intermediate-mass SBs will decay the slowest in these stars (see Appendix~\ref{sec:appendix:numerics}).

\section{Wind tunnel numerical framework}\label{sec:cewt}
Global simulations of planetary engulfment that account for changes in the internal structure of the post-MS star and the SB are computationally challenging and expensive because they involve two vastly different scales. The scale of the flow in the vicinity of the SB is \( {\approx}\max\lp R_\text{SB},R_a\rp\approx R_\text{Jup} \), while the scale of the orbit and the star is \( \approx\qty{1}{\astronomicalunit}\approx10^3R_\text{Jup} \). This disparity of scales motivates isolating the processes occurring at each scale, not only for computational feasibility and accuracy, but also to understand the role of each of these processes and eventually the interplay between the processes at different scales. Here we perform simulations of the flow within a few \( \max\lp R_\text{SB},R_a\rp \) of the SB\@. We aim to understand the morphology of the flow in the vicinity of the SB, and the forces acting on it. Measurements of these forces allow numerical integration of the equation of motion of the SB inside the star. This approach allows inexpensive yet reasonably accurate exploration of a much larger region of parameter space.

\begin{figure}[t!]
    \centering
    \includegraphics[width=\columnwidth]{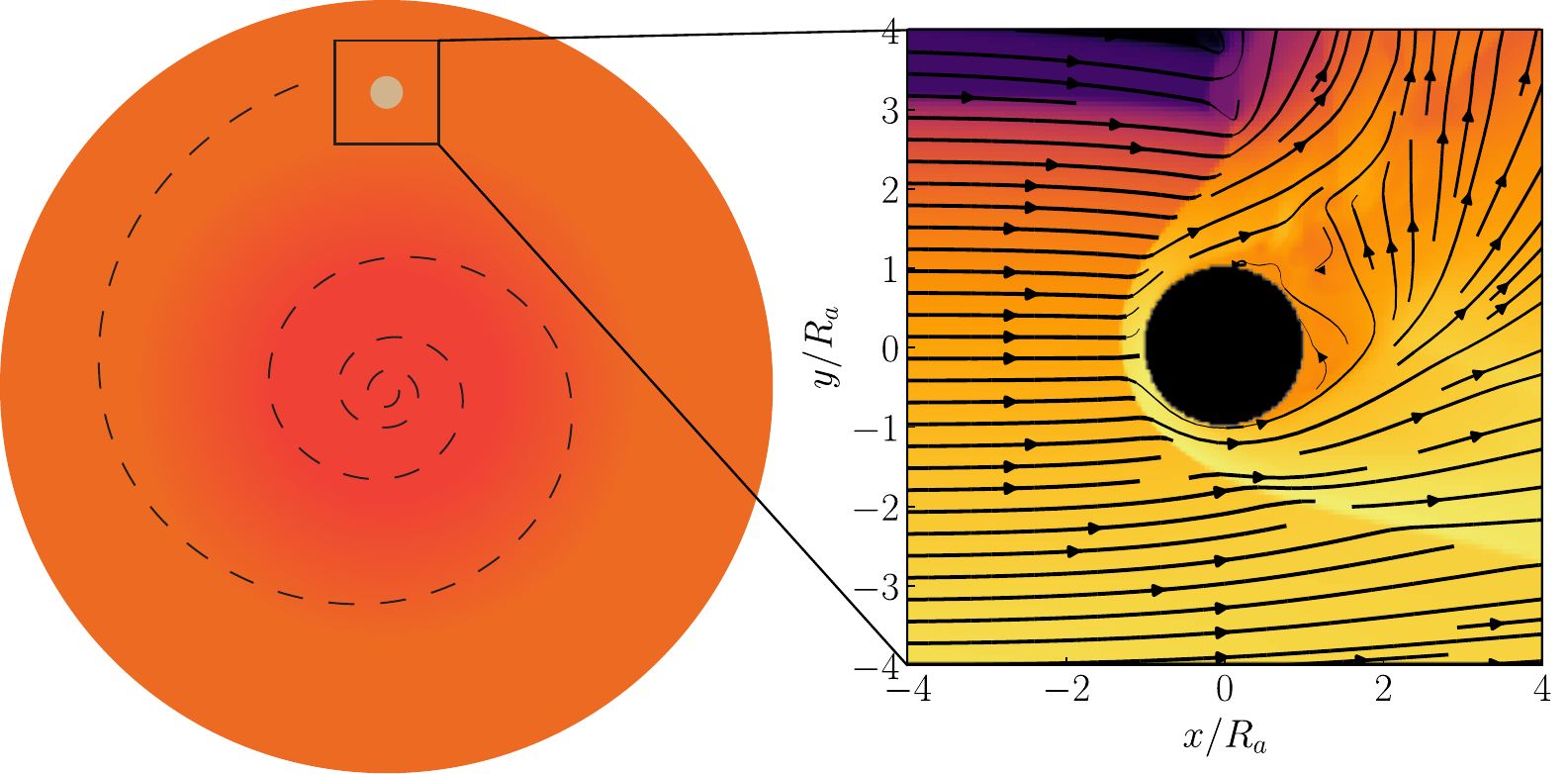}
    \caption{Schematic of the wind tunnel numerical setup. Left: substellar body (SB) embedded in the envelope of a giant star, with an orbital decay trajectory shown as a dashed line. Right: density slice and velocity streamlines in a wind tunnel simulation of the embedded object. The wind velocity and density gradient point in the \( \hat{x} \) and \( -\hat{y} \) directions, respectively. The orbital speed is higher than the local sound speed, so the object will drive a shock as it moves through the stellar envelope.}\label{fig:setup}
    \end{figure}

We use the ``wind tunnel'' numerical setup \citep{MacLeod2017}, illustrated in Figure~\ref{fig:setup}. We simulate a local domain in the frame of the engulfed SB, and we supply a time-independent ``wind'' from the \( -\hat{x} \) direction. The flow morphology (and therefore the drag coefficients) is uniquely determined by a set of dimensionless parameters: the mass ratio \( q\equiv M_\text{SB}/M_\text{enc} \), the dimensionless density gradient \( \varepsilon_\rho \) (equation~\ref{eq:eps_rho}), the mach number \( \mathcal{M} \), and the ratio between geometrical and gravitational radii \( R_\text{SB}/R_a \). The equation of hydrostatic equilibrium for the envelope \citep{MacLeod2017} relates these parameters as
\begin{equation}
    \min\lp1,\frac{R_a}{R_\text{SB}}\rp\varepsilon_\rho = \frac{2 q}{\lp 1 + q \rp ^ 2}f_k^{-4} \mathcal{M}^2,
    \label{eq:hse}
\end{equation}
where \( f_k \) is the ratio between the speed of the companion relative to the surrounding material and the Keplerian speed. Equation~\ref{eq:hse} implies three dimensionless parameters are enough to uniquely specify the flow morphology; for most of this work we use the set \( \lp q, \varepsilon_\rho, R_\text{SB}/R_a\rp \). Since the drag coefficients are a function of only these dimensionless parameters, a single simulation can determine the drag for a variety of physical systems. The drag coefficients do not depend on dimensional quantities that set the scale of the system. These quantities do not change the flow morphology, and the total drag scales with them in known ways (for example, linearly, in the case of the density). For simplicity, we set the density at \( y=0 \) to \( \qty{1}{\gram\per\cubic\centi\meter} \), and the wind speed to \( \qty{1}{\centi\meter\per\second} \). We set the pressure at \( y=0 \) so that the flow has the Mach number implied by hydrostatic equilibrium and the rest of the dimensionless parameters.

After computing the flow properties at \( y=0 \), we integrate the equations of hydrostatic equilibrium for a massless atmosphere up to the boundaries in the \( \pm\hat{y} \) directions. In the \( -\hat{y} \) direction, we extend hydrostatic equilibrium to the ghost zones. We add the gravitational field of the enclosed stellar mass so that the gas remains in hydrostatic equilibrium in the absence of external forces. The external force in our simulations, leading to the deflection of the gas, is the gravity of the SB\@. For all other boundaries, we use outflow boundary conditions, in which gas can leave the domain but not enter it. For more details, see \citet{MacLeod2017}. The numerical setup is publicly available (see Appendix~\ref{sec:appendix:reproducibility}).

\begin{figure*}
\centering
\includegraphics[width=\textwidth]{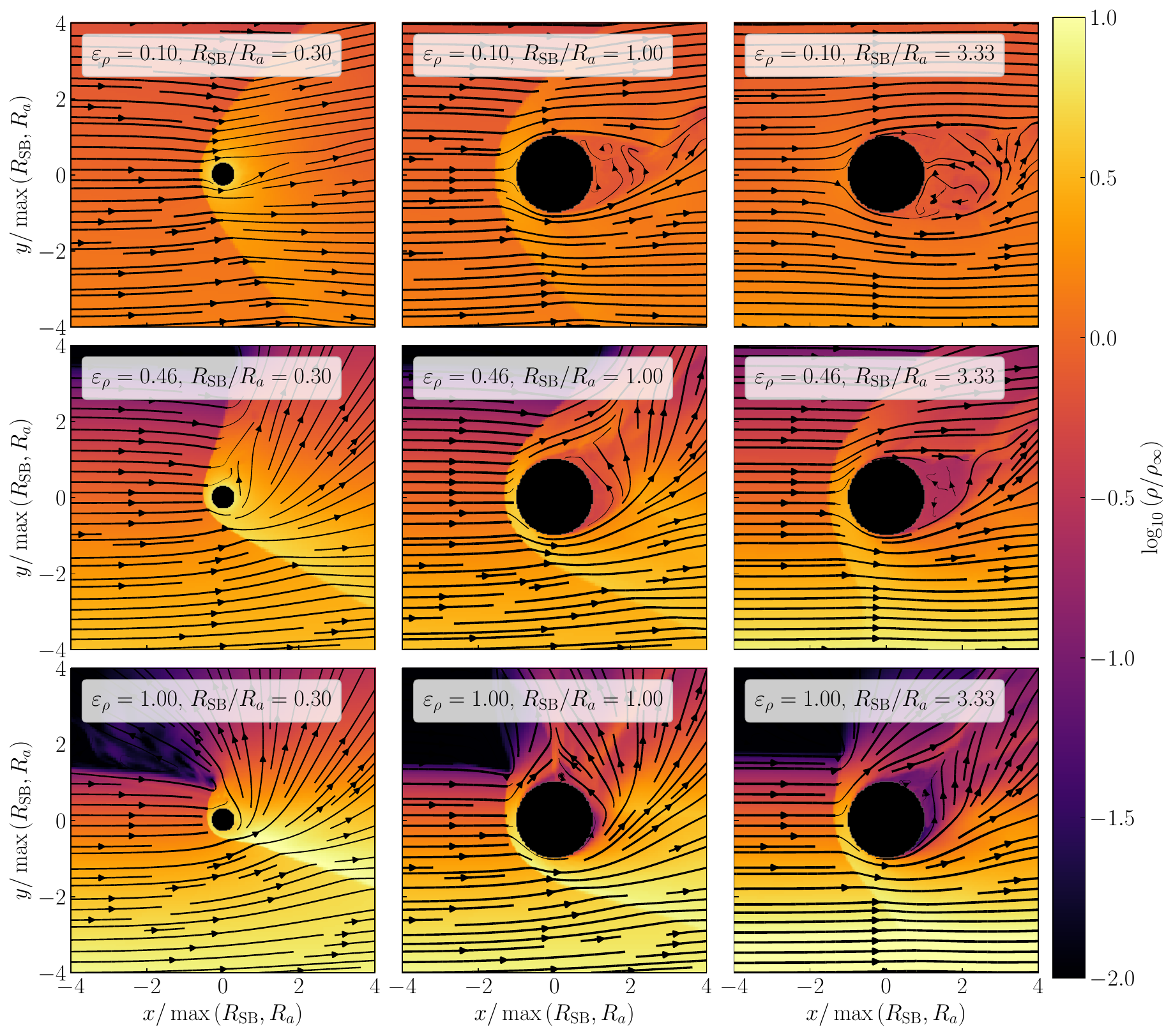}
\caption{Density (in units of upstream density at \( y=0 \), \( \rho_\infty \)) slices in wind tunnel simulations, for several values of the ratio between geometrical and gravitational radii \( R_\text{SB}/R_a \), and the number of density scale heights per \( \max\lp R_\text{SB}, R_a\rp \). These slices are at \( t=20\max\lp R_\text{SB},R_a\rp/v_\text{orb} \), where \( v_\text{orb} \) is the upstream gas speed. At low \( R_\text{SB}/R_a \), a spherically symmetric envelope of material that cannot be accreted forms around the SB, suppressing ram pressure drag. At high \( R_\text{SB}/R_a \), a vacuum forms behind the SB, increasing ram pressure drag and suppressing gravitational drag. An animated version of this figure is available in the HTML version of the article. The animation shows the time evolution of the density slices from \( t=0 \) to \( t=20\max\lp R_\text{SB},R_a\rp/v_\text{orb} \).}\label{fig:wind_tunnel_grid}
\end{figure*}

We set \( f_k=1 \), thereby assuming a circular orbit and no corotation between the star and the SB\@. During the post-main-sequence, synchronization tides shrink the orbit of the companion and transfer its angular momentum into the star. Small SBs do not have enough orbital angular momentum to bring the star into corotation (equality of the orbital and rotational periods of the star), leading to orbital decay through the \cite{Darwin1879} instability. The ratio between rotational and orbital periods at the onset of the engulfment of a Jupiter-like planet by a Sun-like host star evolved to \( 0.05R_\odot \) is \( \approx0.2 \) \citep{Gallet2017}; for gas giant planets engulfed by a \( 1.5M_\odot \) host star, this ratio is \( \lesssim0.3 \) \citep[Table A.1 of][]{Privitera2016a}. More massive SBs, such as brown dwarfs, will enhance stellar rotation more.

In the wind tunnel framework, the density gradient (pointing in the \( -\hat{y} \) direction) and the velocity of the stellar material in the frame of the SB (pointing in the \( \hat{x} \) direction) are perpendicular. This configuration doesn't accurately represent the flow around the SB when its orbit is eccentric. Tides significantly lower the orbital eccentricity of closely orbiting planets throughout stellar evolution, while the eccentricity of distant planets remains roughly constant \citep[see Figure 8 of][]{Villaver2014}. The average eccentricity of the known exoplanets more massive than Jupiter around stars between \( 0.7M_\odot \) and \( 1.5M_\odot \) is \( \approx0.2 \) \citep{exoplanetarchive}. As a result of tides, these planets will likely be engulfed in orbits more circular than their current ones. However, there is theoretical and observational evidence for a transient population of exoplanets orbiting evolved stars at moderate eccentricity \citep{Villaver2014,Grunblatt2022}, whose orbits decay before they circularize.

This framework also neglects the curvature of the velocity field within the domain. This approximation is valid when the half-length of the domain \( L_\text{domain}/2 \) is much smaller than the orbital separation \( r \). We set the domain length to \( 10\max\lp R_\text{SB}, R_a\rp \). In the geometrical regime, the condition \( L_\text{domain}/2<r \) reduces to \( R_\text{SB}/R_a<\lp1+q^{-1}\rp/10 \). In the gravitational regime, it reduces to \( q<1/9 \). Figure~\ref{fig:parameter_space} shows these constraints.

\begin{figure*}
\includegraphics[width=0.495\textwidth]{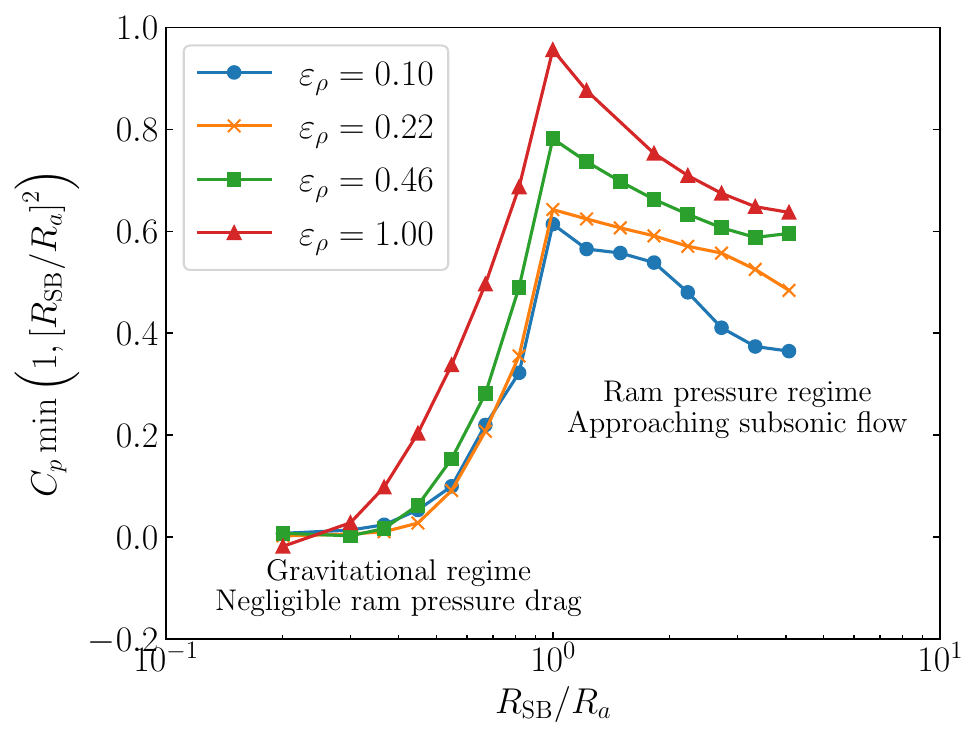}
\includegraphics[width=0.495\textwidth]{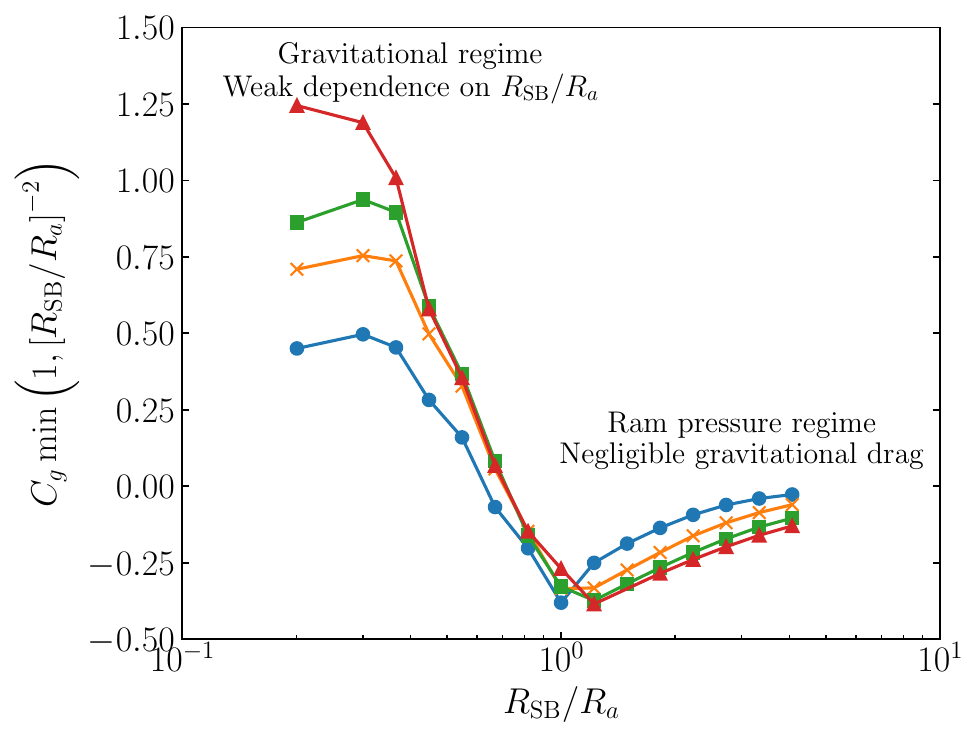}
\caption{Ram pressure (left) and gravitational (right) drag coefficients for a set of simulations with mass ratio \( q=2.15\times10^{-2} \), as a function of the ratio between geometrical and gravitational radii, \( R_\text{SB}/R_a \). Each line corresponds to a different dimensionless gradient \( \varepsilon_\rho \). The drag coefficients depend on \( R_\text{SB}/R_a \) most strongly during the transition regime \( R_\text{SB}\sim R_a \), then have a weaker dependence until the mach number approaches (and goes below) unity.}\label{fig:drag_coefficients}
\end{figure*}%

We wrote this numerical framework as a setup for the FLASH \citep{Fryxell2000,Dubey2014,Dubey2015} code. It uses FLASH to solve the equations of inviscid hydrodynamics on a Cartesian mesh with adaptive mesh refinement. We use an ideal gas equation of state \( P=\lp \gamma - 1 \rp u \), where \( P \) is the pressure, \( \gamma \) the ratio of specific heats (which we take to be \( 5/3 \)), and \( u \) is the internal energy per unit volume. The base resolution of our simulations is \( 160 \) cells per dimension. We use adaptive mesh refinement with a criterion based on proximity to the surface of the SB\@; we choose the maximum level of refinement such that there are always \( \approx60 \) cells across the SB\@. See Appendix~\ref{sec:appendix:numerics:windtunnel} for hydrodynamics convergence tests.

\subsection{Model for the substellar body}
Previous wind tunnel CEE simulations between extended stars and compact objects have modeled the compact object as a numerical ``sink.'' Inside the sink, the simulation multiplies fluid variables by a small number, creating a numerical vacuum that emulates accretion of the surrounding material onto the object. In real systems, these objects accrete because the material accumulating around them is hot and dense enough to cool via neutrino emission \citep{MacLeod2015a,Fragos2019}. There is no such cooling channel for material near the surface of the SB, and the gas is too optically thick to cool radiatively. The timescale over which radiation will carry energy through the optically thick surrounding material, allowing it to be accreted, is much longer than the orbital decay timescale. We therefore model the SB as a rigid body with a reflective surface. We use FLASH's unsplit hydrodynamics solver, which is based on the unsplit magnetohydrodynamics solver \citep{Lee2013}. The code applies the reflective boundary condition at the surface of the rigid body, i.e., in the rigid body cells adjacent to the fluid cells. The reflective boundary implementation in FLASH requires a Courant--Friedrichs--Lewy (CFL) number \( \leq0.3 \) for numerical stability because the reconstruction order near the boundary is lower.

During the engulfment of a compact object by a giant star, the geometrical size of the object is negligible (\( R_\text{SB}\ll R_a \)), whereas an engulfed SB can have \( R_\text{SB}\gtrsim R_a \). When the radius of the ``sink'' object is \( \gtrsim0.2R_a \), the shock morphology changes into a ``tail shock'' that trails the path of the object \citep[Figure 10 in][]{Ruffert1994}. As we will see (Figure~\ref{fig:wind_tunnel_grid}), the reflective boundary prevents this change in morphology. These qualitative differences motivate modified wind tunnel simulations that more accurately represent how engulfed SBs interact with their surroundings, and the associated flow morphology and drag coefficients.

We do not model the internal structure of the SB, whose mass loss and deformation could affect the flow morphology around it, and the cross-section for interactions with its surroundings. Ram pressure will flatten the surface of the SB facing the incoming flow and compress the SB, making it harder to disrupt \citep{Jia2018}. Gradual ablation of the SB is unlikely to destroy the SB before ``global'' processes that act on its dynamical timescale \citep{Passy2012,Jia2018}, although the hydrodynamics of ablation in this context are uncertain. We discussed these global processes in Section~\ref{sec:landscape:destruction}; we evaluate their corresponding destruction criteria using only the unperturbed SB structure.

\subsection{Drag force measurements}\label{sec:methods:forces}
We measure the forces on the object when the simulation reaches steady state, which takes a few flow-crossing times \( \tau_\text{c}=\max\lp R_a, R_\text{SB}\rp/v_\text{orb} \). We measure the gravitational drag force by integrating the gravitational force of the surrounding density field up until \( 1.6\max\lp R_\text{SB}, R_a\rp \), as in \cite{MacLeod2017}. The ram pressure drag force is
\begin{equation}
    \mathbf{F}_\text{p} = - \oint_S P \, d\mathbf{A},
\end{equation}
where \( d\mathbf{A} \) is the area element and \( S \) is the surface of the SB\@. Our setup uses FLASH to solve the equations of inviscid hydrodynamics; while the discretization of the equations results in numerical viscosity, there is not a boundary layer around the SB, which could change the ram pressure drag and the viscous stresses acting on it. For the same reason we do not study the dependence of the ram pressure drag on the Reynolds number.

The coefficients measured from the steady state simulations are valid if the timescale over which the simulation reaches steady state is much shorter than the orbital decay timescale. For SBs dominated by ram pressure, whose orbital decay time is equation~\ref{eq:t_decay_ram}, \( \tau_c/\tau_\text{insp,p} = \rho / \langle\rho_\text{SB}\rangle < \langle\rho_\text{enc}\rangle / \langle\rho_\text{SB}\rangle \), where the inequality holds because the density decreases monotonically with radius. This ratio is less than unity at the onset of engulfment; when it reaches unity, the core tidally disrupts the SB (equation~\ref{eq:disruption:tides}). For SBs dominated by gravitational drag, \( \tau_c / \tau_\text{insp,g} = \lp M_\text{SB}/M_\text{enc}\rp^2 \rho / \langle\rho_\text{enc}\rangle<1 \).

\begin{figure}
\includegraphics[width=\columnwidth]{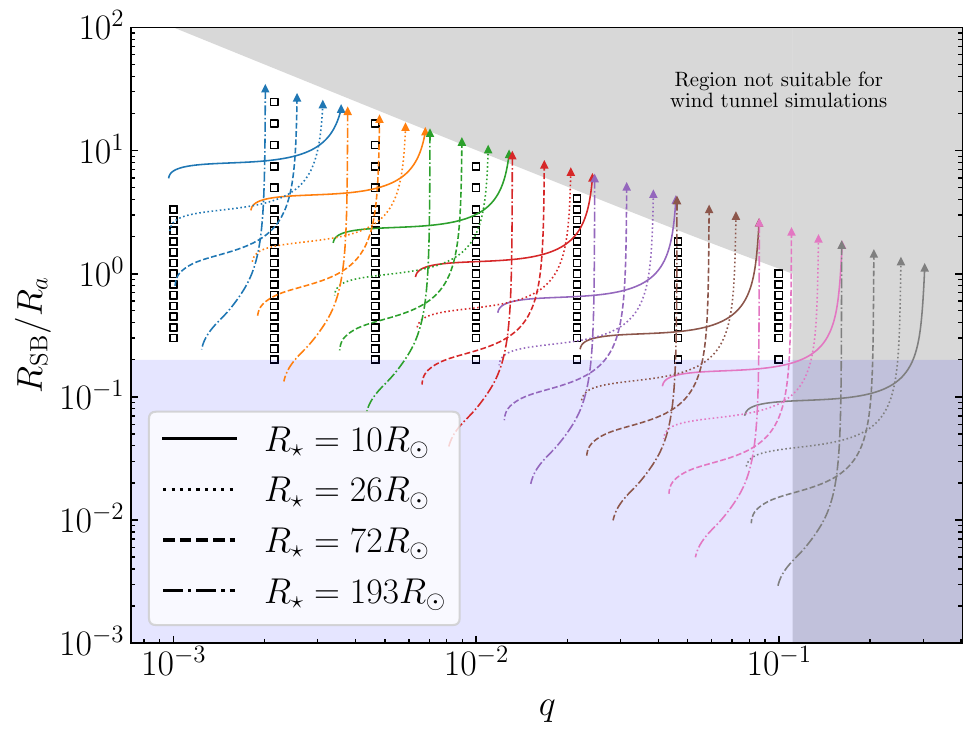}
\includegraphics[width=\columnwidth]{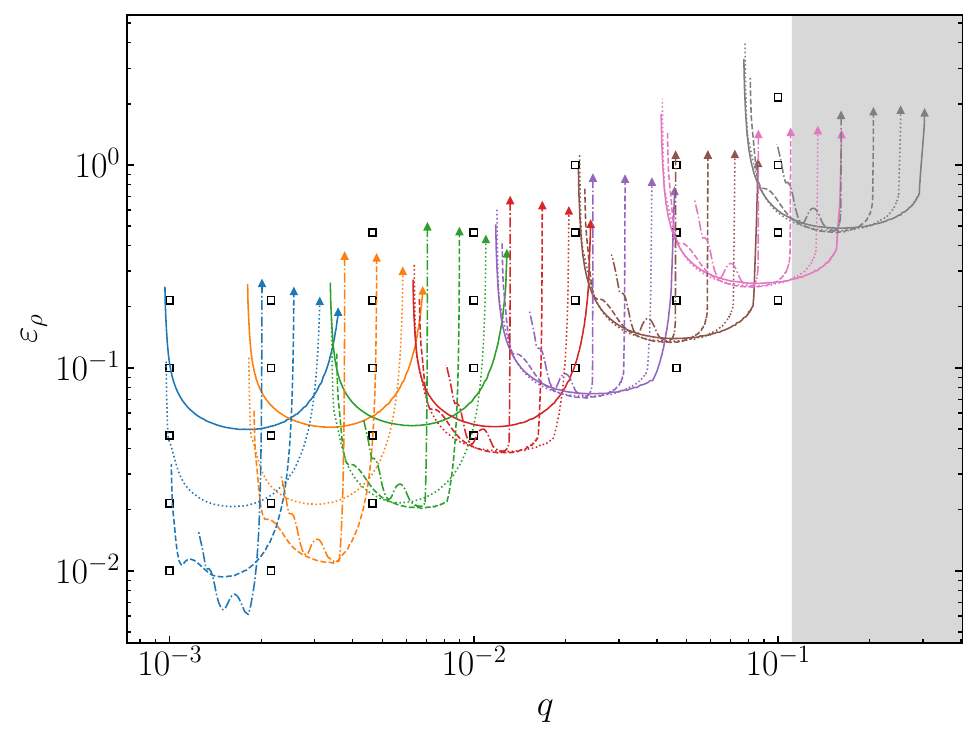}
\includegraphics[width=\columnwidth]{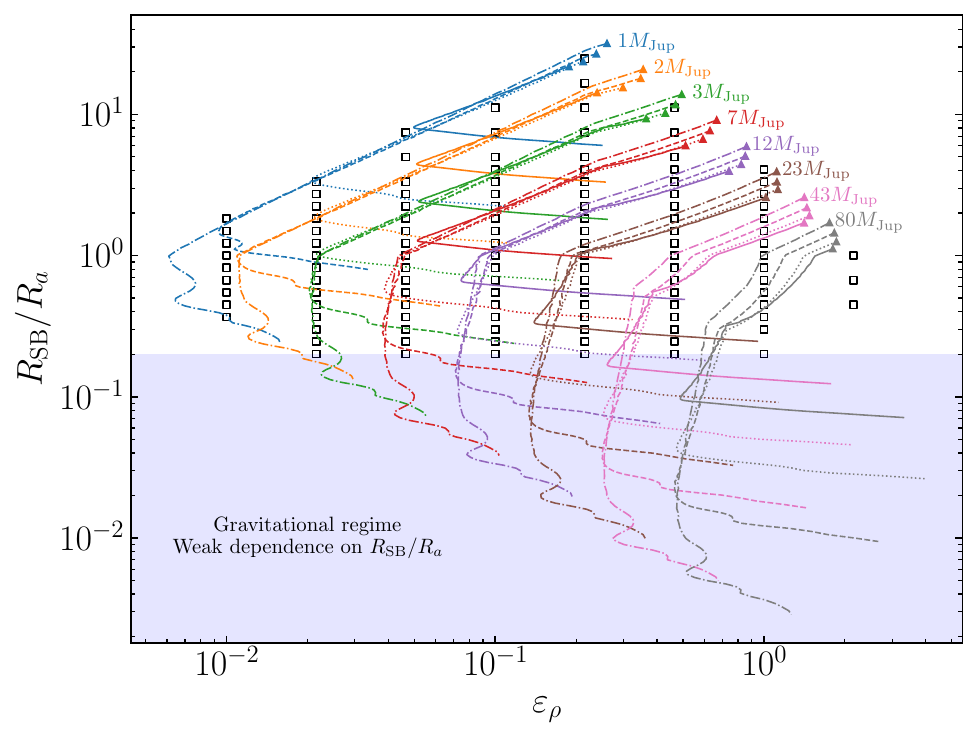}
\caption{Trajectories in parameter space followed by substellar bodies (SBs) of varying masses after being engulfed by a \( 1M_\odot \) star at varying evolutionary stages. Each line style represents a different evolutionary stage (see legend on top panel), and each color represents an SB of a different mass (see inline labels on bottom panel). Each line starts at a radial coordinate \( r=0.9R_\star \), and ends at the point of SB destruction, represented by a triangle. The squares represent the hydrodynamical simulations done in this work.}\label{fig:parameter_space}
\end{figure}

\section{Flow morphology and drag}\label{sec:morphology}
\subsection{The gravitational and geometrical regimes}\label{sec:regimes}

Figure~\ref{fig:wind_tunnel_grid} shows steady state density slices of wind tunnel simulations with \( q=2.15\times10^{-2} \), at several \( R_\text{SB}/R_a \) and \( \varepsilon_\rho \). At low \( R_\text{SB}/R_a \) (the leftmost column in the Figure), the SB gravitationally focuses gas as in the Bondi--Hoyle--Lyttleton formalism. However, since the SB cannot accrete, this focused material accumulates at its surface. The pressure force exerted by this material opposes the pressure force from material accumulated in front of the object a result of compression when the SB moves through the gas. The resulting pressure field at the surface of the object is spherically symmetric, suppressing ram pressure drag \citep{Thun2016}. As \( R_\text{SB}/R_a \) decreases, ram pressure drag becomes less important not only because the geometrical cross-section decreases, but also because of morphological changes to the flow.
\begin{figure*}
    \centering
        \includegraphics[width=\textwidth]{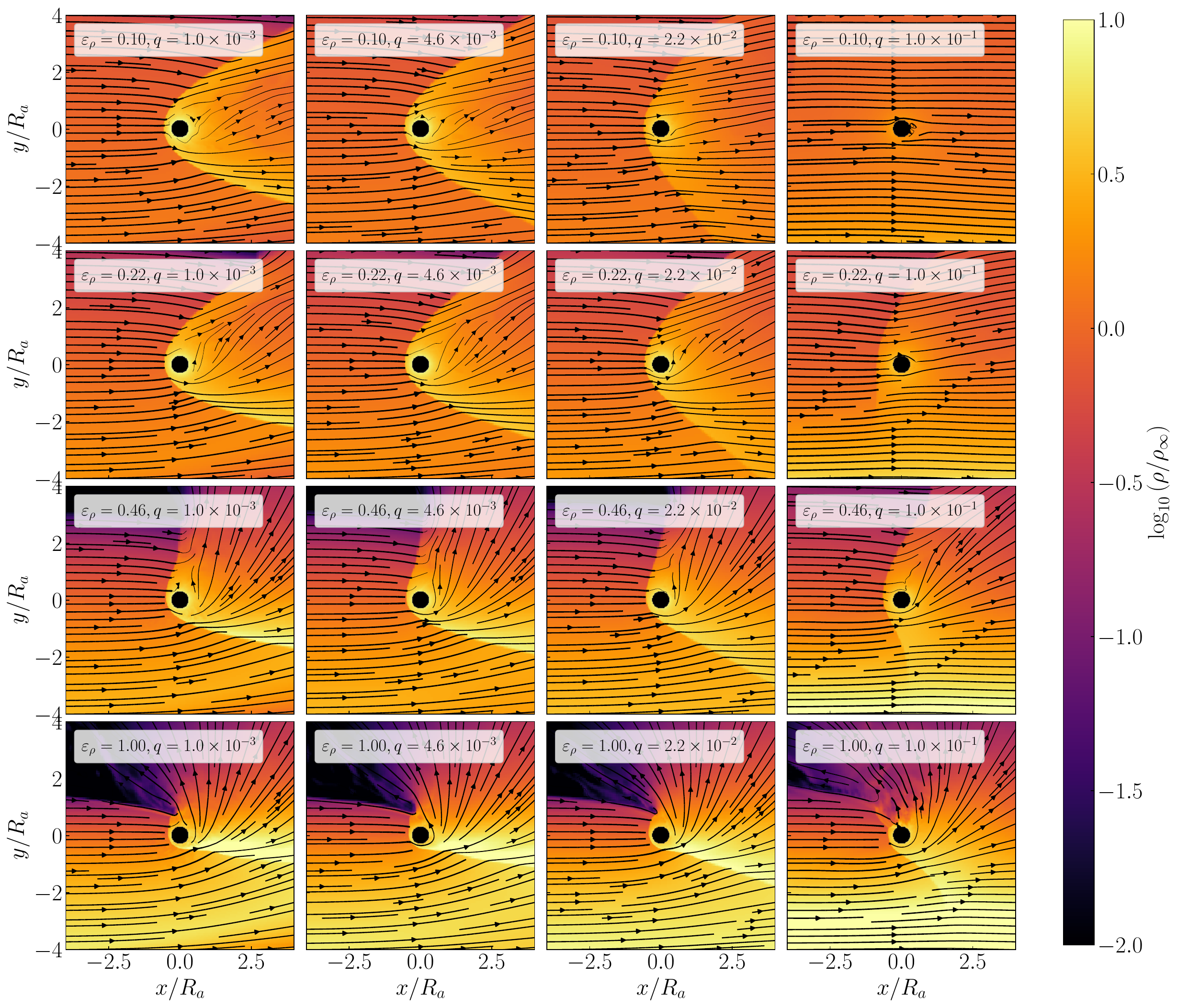}
    \caption{Density (in units of upstream density at \( y=0 \), \( \rho_\infty \)) in wind tunnel simulations as a function of space. Each panel shows a different combination of the dimensionless density gradient (\( \varepsilon_\rho \)), and the ratio between the mass of the substellar object and the mass enclosed by its orbit (\( q \)). The ratio between geometrical and gravitational radii is \( 0.3 \) in all panels. These slices are at \( t=20R_a/v_\text{orb} \), where \( v_\text{orb} \) is the upstream gas speed. Larger density gradients and Mach numbers result in higher drag coefficients. An animated version of this figure is available in the HTML version of the article. The animation shows the time evolution of the density slices from \( t=0 \) to \( t=20R_a/v_\text{orb} \).}\label{fig:morph}
\end{figure*}%

When \( R_\text{SB}\sim R_a \), the gravity of the SB is not strong enough to deflect the surrounding material, and a low-density region forms behind the SB\@. The ram pressure force exerted by gas in front of the SB is now unopposed, and gravitational drag decreases significantly. When \( R_\text{SB}\gg R_a \), the material in front of the SB dominates gravitational drag (as opposed to material behind the SB, as when \( R_\text{SB}\ll R_a \)), and the gravitational drag coefficient becomes negative. Similar results have been found for compact objects with outflows \citep{Gruzinov2020,Li2020,Kaaz2022} and luminous planetesimals moving through a disk \citep{Masset2017a,Masset2017b}. In those settings, feedback from the object can interfere with the flow at impact parameters \( \sim R_a \) that the SB would have gravitationally focused had there been no feedback. In our simulations, the rigidity of the SB and its reflective surface have this effect when \( R_\text{SB}>R_a \).

Figure~\ref{fig:drag_coefficients} shows the drag coefficients for the same set of simulation parameters as Figure~\ref{fig:wind_tunnel_grid}. In the gravitational regime, the drag force is most naturally written as
\begin{equation}
    \label{eq:drag_grav_dominated}
    \mathbf{F}_\text{drag}=-\pi\rho v^2R_\text{a}^2\ls C_g + \lp\frac{R_\text{SB}}{R_a}\rp^{2}C_p\rs\hat{v},
\end{equation}
where \( \hat{v} \) is the unit vector in the direction of the SB's velocity. On the other hand, in the geometrical regime
\begin{equation}
    \label{eq:drag_ram_dominated}
    \mathbf{F}_\text{drag}=-\pi\rho v^2R_\text{SB}^2\ls C_p + \lp\frac{R_\text{SB}}{R_a}\rp^{-2}C_g\rs\hat{v}.
\end{equation}
Equations \ref{eq:drag_grav_dominated} and \ref{eq:drag_ram_dominated} are equivalent, but it is often convenient to write the drag force in the functional form of the dominant source of drag, with a small correction term that accounts for the nondominant source of drag. The drag can be written as a combination of the previous two expressions,
\begin{equation}
    \mathbf{F}_\text{drag}=-\pi\rho v^2 \max\lp R_\text{SB}, R_a \rp^2\lp C_{p,\text{eff}} + C_{g,\text{eff}}\rp\hat{v},
\end{equation}
where
\begin{gather}
    C_{p,\text{eff}} \equiv C_p \min\lp 1, \ls\frac{R_\text{SB}}{R_a}\rs^2 \rp,\\
    C_{g,\text{eff}} \equiv C_g \min\lp 1, \ls\frac{R_a}{R_\text{SB}}\rs^2\rp.
\end{gather}
The primary motivation for writing the drag force this way is that the ratio between the ``effective'' coefficients equals the ratio between the drag forces. Additionally, when the flow converges to the geometrical regime, the effective gravitational drag coefficient approaches zero. The gravitational drag coefficient alone does not approach zero when \( R_\text{SB}\gg R_a \) because in that limit the material in front of the SB dominates gravitational drag. Since larger SBs have more material in front of them, the gravitational force exerted by that material is larger. However, this gravitational drag force increases with \( R_\text{SB}/R_a \) slower than quadratically, so gravitational drag becomes a smaller fraction of the total drag as \( R_\text{SB}/R_a \) increases.

Figure~\ref{fig:drag_coefficients} shows that when \( R_\text{SB}\lesssim0.25 \) the effective drag coefficients are independent of \( R_\text{SB}/R_a \), so that
\begin{gather}
    C_{g,\text{eff}} = C_g = C_g\lp q, \varepsilon_\rho \rp,\\
    C_{p,\text{eff}} = C_p \lp R_\text{SB}/R_a \rp^2 \approx 0.
\end{gather}

On the other hand, the ram pressure drag coefficient depends on \( R_\text{SB}/R_a \) even when \( R_\text{SB}\gg R_a \). This dependence can be understood from the equation of hydrostatic equilibrium, which when \( R_\text{SB}>R_a \) reduces to
\begin{equation}
    \frac{R_a}{R_\text{SB}}\varepsilon_\rho = \frac{2 q}{\lp 1 + q \rp ^ 2}f_k^{-4} \mathcal{M}^2.
\end{equation}
At a fixed mass ratio \( q \) and dimensionless density gradient \( \varepsilon_\rho \), the mach number \( \mathcal{M}\propto\lp R_\text{SB}/R_a\rp^{-1/2} \). As \( R_\text{SB}/R_a \) increases, the mach number decreases, so the density discontinuity across the shock is smaller, reducing ram pressure drag. This gradual transition towards subsonic flow can be seen in the top row of Figure~\ref{fig:wind_tunnel_grid}, from left to right. In the geometrical regime,
\begin{gather}
    C_{g,\text{eff}} = C_g \lp R_a/R_\text{SB} \rp^2 \approx 0,\\
    C_{p,\text{eff}} = C_p = C_p \lp q, \varepsilon_\rho, R_\text{SB}/R_a \rp.
\end{gather}

\subsection{Dependence on mass ratio and dimensionless density gradient}
The dependence of the drag coefficients on the density gradient, the Mach number, and the mass ratio can be understood from the relationship between these parameters in hydrostatic equilibrium (Equation~\ref{eq:hse}), and the flow morphology in Figure~\ref{fig:morph}. Simulations with stronger density gradients have larger drag coefficients because the SB interacts with higher density gas (from deeper in the envelope) both geometrically and gravitationally. From hydrostatic equilibrium, increasing the mass ratio at a fixed Mach number will result in a stronger density gradient, and larger drag coefficients.

Similarly, at a given density gradient, larger Mach numbers result in a narrower shock opening angle, allowing focused material to accumulate closer to \( y=0 \), increasing the horizontal component of the drag force and therefore the gravitational drag coefficient. Higher Mach numbers also result in increased gas compression in front of the SB, increasing the ram pressure drag coefficient. At fixed density gradient, increasing the Mach number requires decreasing the mass ratio, so increasing mass ratios reduce the drag coefficient.

\begin{figure}[t!]
    \centering
    \includegraphics[width=0.49\columnwidth]{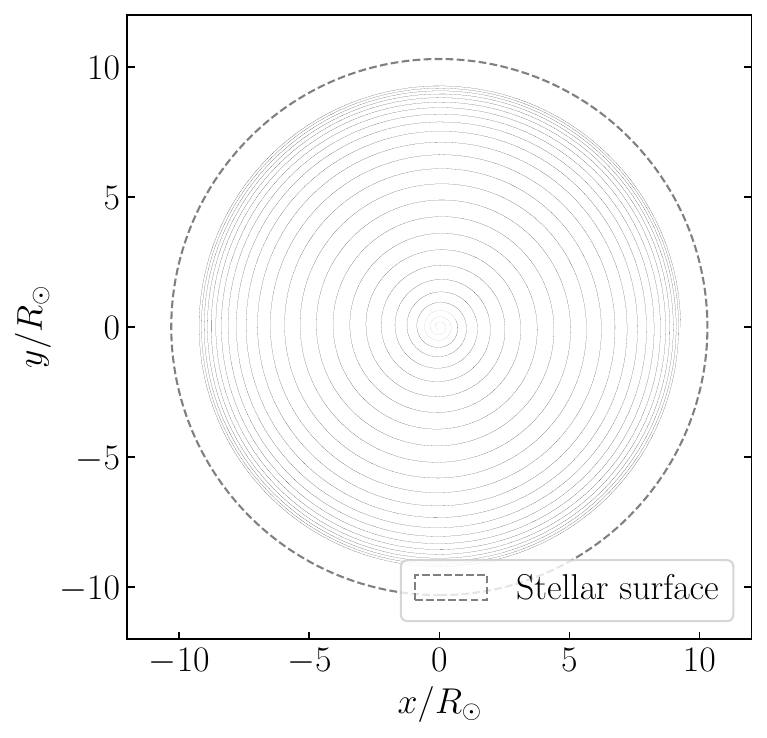}
    \includegraphics[width=0.49\columnwidth]{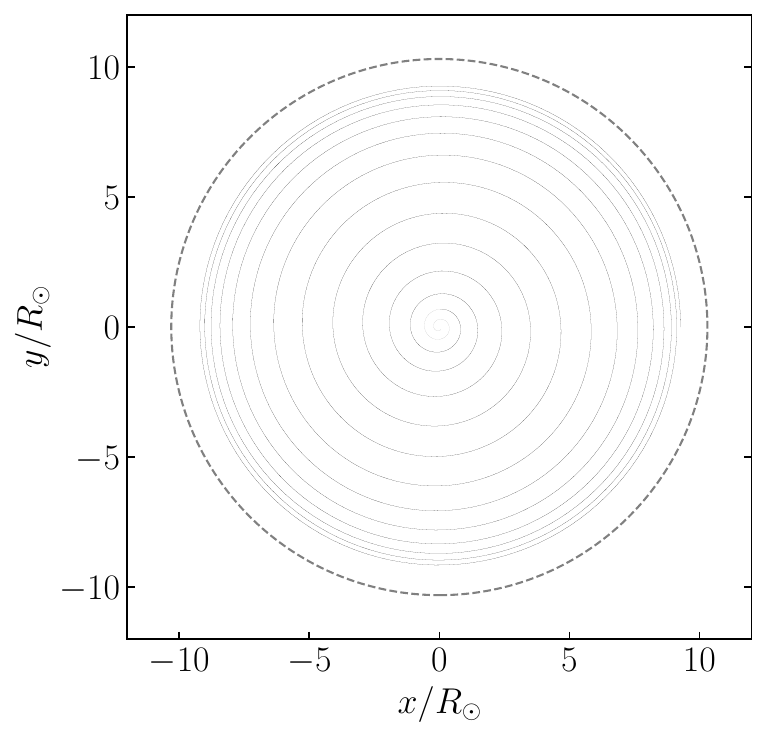}
    \caption{Orbital decay trajectory of a \( 1M_\text{Jup} \) planet inside a \( 1M_\odot \) star evolved to \( 10R_\odot \), as computed using analytical (left panel) and numerical (right panel) drag coefficients. Trajectories turn transparent at the estimated point of destruction. Orbital decay takes \( \approx\qty{24}{\day} \) (a factor of \( \approx2 \) faster) when using numerical drag coefficients.}\label{fig:compare_trajectories_planet}
\end{figure}

\section{Applications to planetary engulfment}\label{sec:applications}
\subsection{Parameter space for hydrodynamical simulations}
Figure~\ref{fig:parameter_space} shows the trajectories of engulfed SBs in parameter space, for several SB masses and stellar evolutionary stages. Each line style corresponds to a stellar evolutionary stage, and each color corresponds to an SB mass. Each line starts at a radial coordinate \( r=0.9R_\star \) and ends at the point of SB destruction, represented by a triangle. This figure, as Figure~\ref{fig:flow_params}, assumes SBs are in circular orbits throughout engulfment. The wind tunnel framework cannot simulate the region in parameter space where the domain size would be comparable to the orbital separation (the gray region in the top and middle panels). We discussed this limitation quantitatively in Section~\ref{sec:cewt}.

\begin{figure}[t!]
    \centering
    \includegraphics[width=0.49\columnwidth]{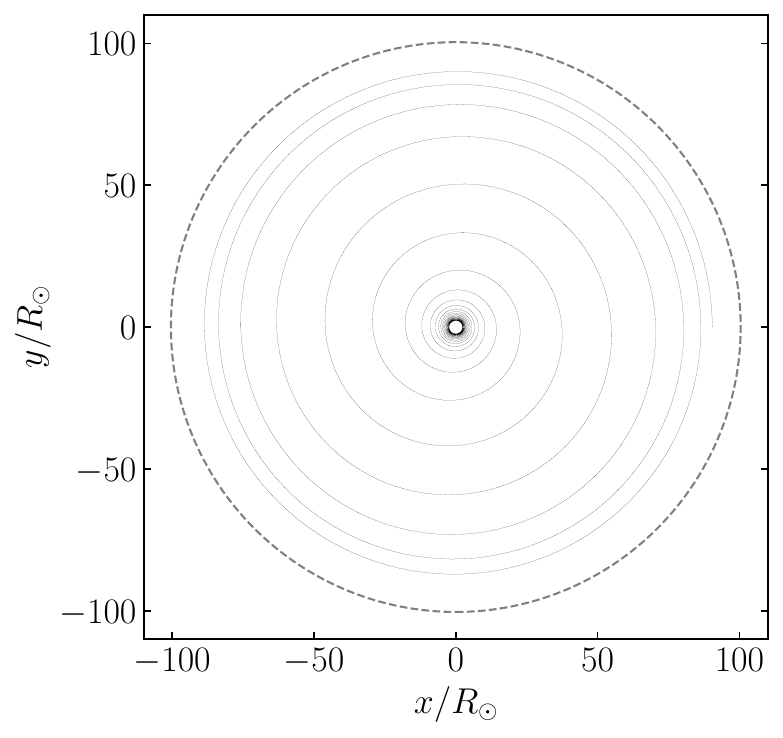}
    \includegraphics[width=0.49\columnwidth]{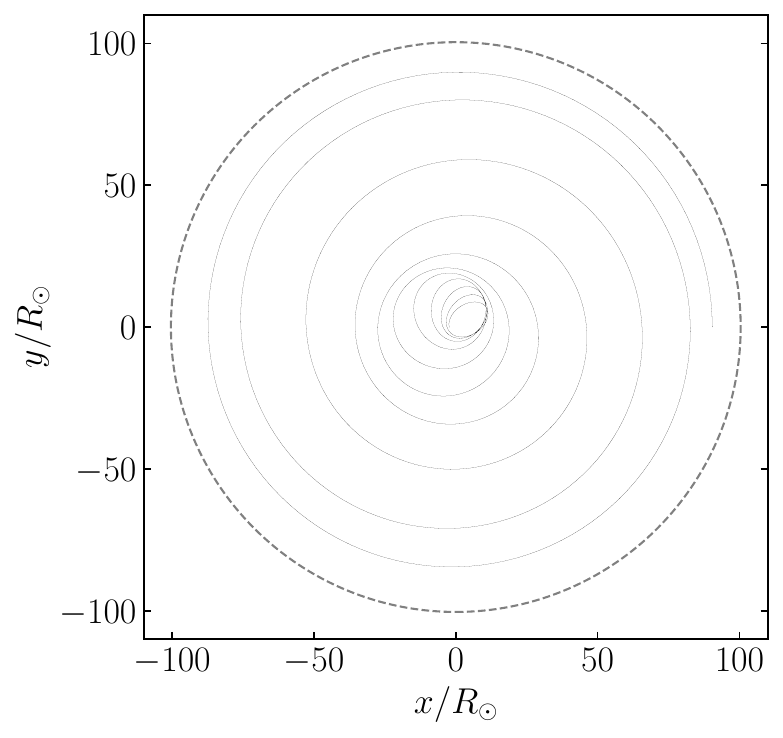}
    \caption{Orbital decay trajectory of a \( 80M_\text{Jup} \) brown dwarf inside a \( 1M_\odot \) star evolved to \( 100R_\odot \), as computed using analytical (left panel) and numerical (right panel) drag coefficients. For readability, we show the trajectories only down to an orbital separation of \( 2.5R_\odot \). While the orbital decay timescale is similar in both cases (\( \approx\qty{1.3}{\year} \)), numerical drag coefficients result in a more eccentric orbit.}\label{fig:compare_trajectories_brown_dwarf}
\end{figure}

The top panel shows that more massive SBs (those with higher \( q \)) are deeper in the gravitational regime, and that \( R_\text{SB}/R_a \) increases throughout engulfment. The destruction points for SBs with \( q\gtrsim5\times10^{-3} \) lie in a line; these SBs are destroyed by tidal disruption. From the definition of the tidal radius, tidal disruption will occur when the orbital separation \( r=R_\text{SB}q^{-1/3} \). Under the assumption of circular orbits, \( R_a=2q r \). Combining these two equations, \( R_\text{SB}/R_a=q^{-2/3}/2 \) at the point of destruction, as seen in the figure.

This figure determines the parameter space that we must simulate to capture the diverse flow morphologies of the systems that undergo engulfment. In Section~\ref{sec:regimes}, we found that the drag coefficients are independent of \( R_\text{SB}/R_a \) in the gravitational regime. The blue region in Figure~\ref{fig:parameter_space} shows this regime. Therefore, hydrodynamical simulations need to span only the transition and geometrical regimes. Each square in the Figure represents a hydrodynamical simulation, of which we ran 428.

\begin{figure*}[ht!]
    \centering
    \includegraphics[width=\textwidth]{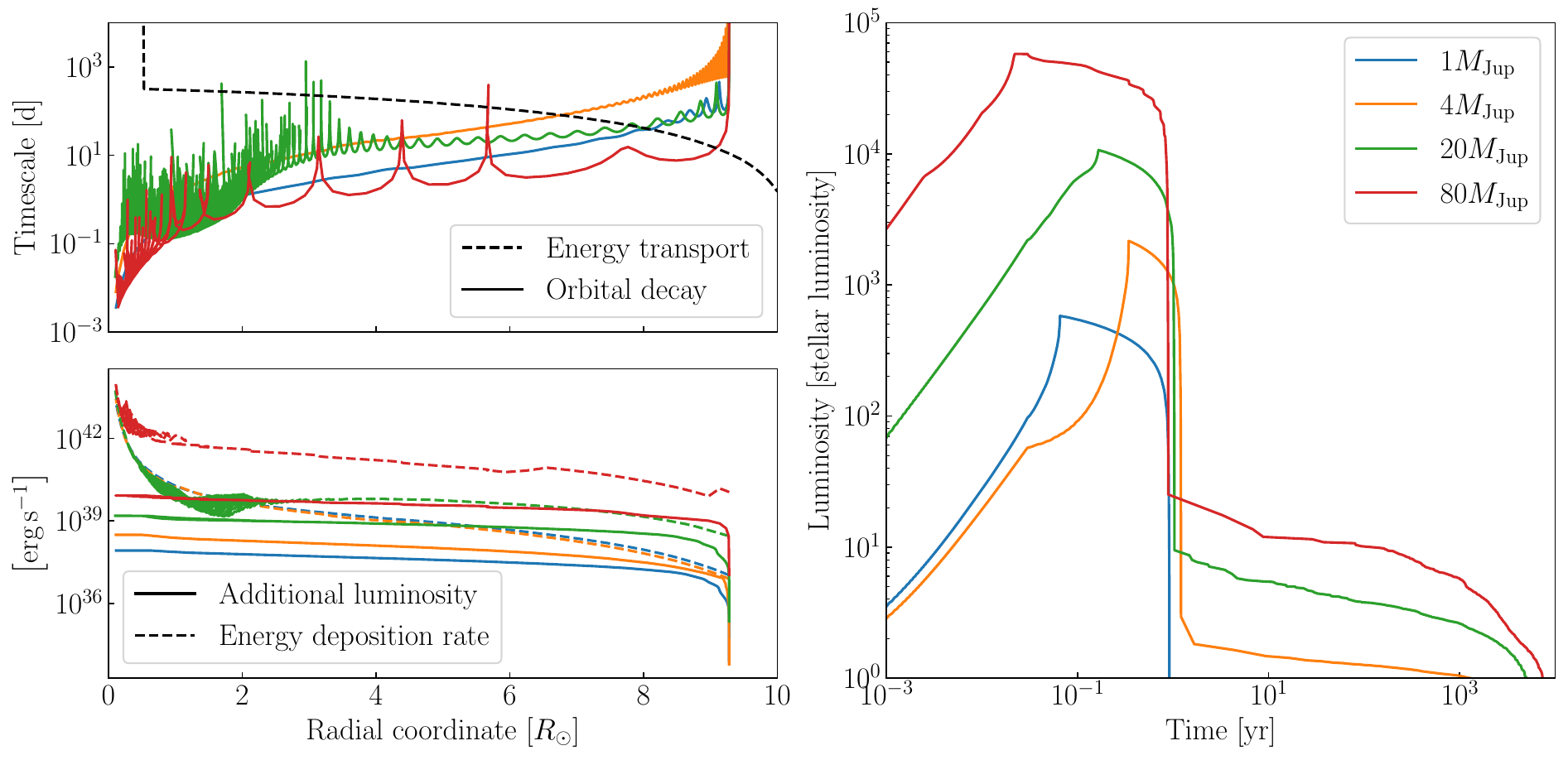}
    \caption{Top left panel: orbital decay and energy transport timescales as a function of the radial coordinate \( r \) of the substellar body (SB) inside the star. The orbital decay timescales for the two more massive substellar bodies show peaks because we define the orbital decay timescale as \( \labs r/\dot{r}\rabs \), and as a result of eccentricity in the orbit \( \dot{r} \) changes sign. Bottom left panel: rate of energy deposition and luminosity (excluding the stellar luminosity), again as a function of position inside the profile. When the energy transport timescale time is much longer than the orbital decay timescale, the luminosity is much smaller than the energy deposition rate. Right panel: luminosity as a function of time, in units of the unperturbed stellar luminosity.}\label{fig:luminosity}
\end{figure*}

\subsection{Orbital decay trajectories}\label{sec:methods:inspiral}
The equation of motion for the SB inside the star is
\begin{equation}
    M_\text{SB}\frac{d\mathbf{v}}{dt}=    \lp M_\text{SB}-\rho V_\text{SB}\rp\mathbf{g}
    +\mathbf{F}_\text{drag},
    \label{eq:eom}
\end{equation}
where \(\mathbf{v}\) is the velocity of the SB, \(t\) is time, \(V_\text{SB}\) is the volume of the SB, and \(\mathbf{g}=-G M_\text{enc}\mathbf{\hat{r}}/r^2\) is the gravitational acceleration from the mass enclosed by the SB's orbit. The term \( \rho V_\text{SB}\mathbf{\hat{g}} \) is the buoyancy acting on the SB, which is important when the local density equals the average density of the SB\@. Since the local density is always smaller than the average enclosed density, and the SB will be tidally disrupted approximately when the average enclosed density is equal to its own mean density, buoyancy won't be important before SB destruction.

We integrate equation~\ref{eq:eom} numerically using the IAS15 \citep{Rein2015} integrator from the N-body code REBOUND \citep{Rein2012}. At every timestep, we compute the dimensionless parameters ($q$, $\varepsilon_\rho$, $R_\text{SB}/R_a$) and linearly interpolate the drag coefficients \( C_g \) and \( C_p \) we measured in the hydrodynamical simulations. For points outside the domain of the interpolation, we used the nearest available point.

We initially place the SB in a circular orbit at an orbital separation $0.9R_\star$. At every timestep we interpolate the properties of the stellar profile using the GSL implementation of the Steffen method \citep{Steffen1990}. We stop the integration when \( r\leq R_\text{SB} \), and apply the destruction criteria during postprocessing.

We compute stellar profiles using MESA\@. The orbital decay timescale is \( \lesssim\qty{400}{\year} \) for the systems we consider here (a \( 1M_\text{Jup} \) planet inside a \( 1M_\odot \) star at the tip of the RGB takes \( \approx\qty{400}{\year} \)). On the other hand, the timescales \( R/\dot{R} \) and \( L/\dot{L} \) over which radius and luminosity change significantly, respectively, are \( \gtrsim\qty{e6}{\year} \) throughout the RGB\@. Therefore, stellar structure doesn't change significantly as a result of stellar evolution over the timescales relevant to engulfment. We do not model the effects of engulfment on stellar structure, and use a single unperturbed MESA profile throughout our integration of the equation of motion of the SB\@.

Figure~\ref{fig:compare_trajectories_planet} and Figure~\ref{fig:compare_trajectories_brown_dwarf} contrast the orbital decay trajectories obtained by using either analytical (\( C_p = 0.25 \), \( C_g = 1 \)) or numerical drag coefficients, for two different systems. We chose systems whose orbital decay timescales were the smallest compared to the orbital period at their initial separations, so that their trajectories were easy to visualize. Figure~\ref{fig:compare_trajectories_planet} shows the trajectories for a Jupiter-like planet inside a \( 1M_\odot \) star evolved to \( 10R_\odot \). The orbit remains nearly circular, so the energy deposition profiles are similar when using analytical or numerical drag coefficients. However, the orbit decays a factor \( \approx2 \) faster with the numerical coefficients. The timescale of energy transfer from the orbit into the star determines whether the star will transport the energy to the surface, or react dynamically. This distinction is particularly relevant for the stellar envelope, in which convection can carry energy to the surface quickly, lowering the efficiency of energy deposition \citep{Wilson2019, Wilson2020, Wilson2022}.

Figure~\ref{fig:compare_trajectories_brown_dwarf} shows the trajectories for a \( 80M_\text{Jup} \) brown dwarf inside a star evolved to \( 100R_\odot \). For this system, both models for the drag coefficients yield similar orbital decay timescales, but the orbit is significantly more eccentric when using numerical drag coefficients. As discussed in Section~\ref{sec:cewt}, our hydrodynamical simulations neglect eccentricity, and therefore do not capture the flow morphology when the orbit of the SB is significantly eccentric. However, they show that at least small eccentricities arise from initially circular orbits faster than suggested by analytical drag. The evolution of the eccentricity is important for many types of systems. For the most massive SBs that can eject the envelope, it determines the eccentricity of their orbit around the stellar remnant \citep{Szoelgyen2022}. For small SBs, the eccentricity at the point of destruction determines the redistribution of their enriched material throughout the star.

\subsection{Stellar brightening}\label{sec:luminosity}
Planetary engulfment results in a transient dominated by recombination in mass ejected from the outer layers of the star \citep{Metzger2012}. Since we do not model changes to stellar structure, we study only the long-term emission resulting from the eventual transport of orbital energy from the deep layers of the star to its surface. If an energy \( dE \) is deposited into the star, the upper bound (assuming all energy leaves as radiation) for the average increase in stellar luminosity is \( dL=dE/\tau_\text{KH} \), where \( \tau_\text{KH} \) is the energy transport timescale from the location of energy deposition to the surface. More generally, for continuous energy deposition, the time-averaged additional luminosity at time \( t \) is
\begin{equation}
    \Delta L\lp t\rp = \int_{t_0}^t \frac{\dot{E}}{\tau_\text{KH}}\, dt',
    \label{eq:lum}
\end{equation}
where we determine \( t_0 \) by noticing that energy deposited a time \( t-t_0 \) ago contributes to the increase in the average luminosity only if \( t-t_0\leq \tau_\text{KH}\lp t_0 \rp \), since if \( t-t_0> \tau_\text{KH}\lp t_0 \rp \), the energy deposited at \( t_0 \) has already been radiated away. If the energy deposition increases sharply in the inner regions of the star, \( \dot{E}\approx\delta\lp t'-t\rp \Delta E\lp t'\rp \) in equation~\ref{eq:lum}, and \( \Delta L\approx \Delta E/\tau_\text{KH} \) \citep[e.g.,][]{MacLeod2018}.

\begin{figure}[t!]
    \centering
    \includegraphics[width=\columnwidth]{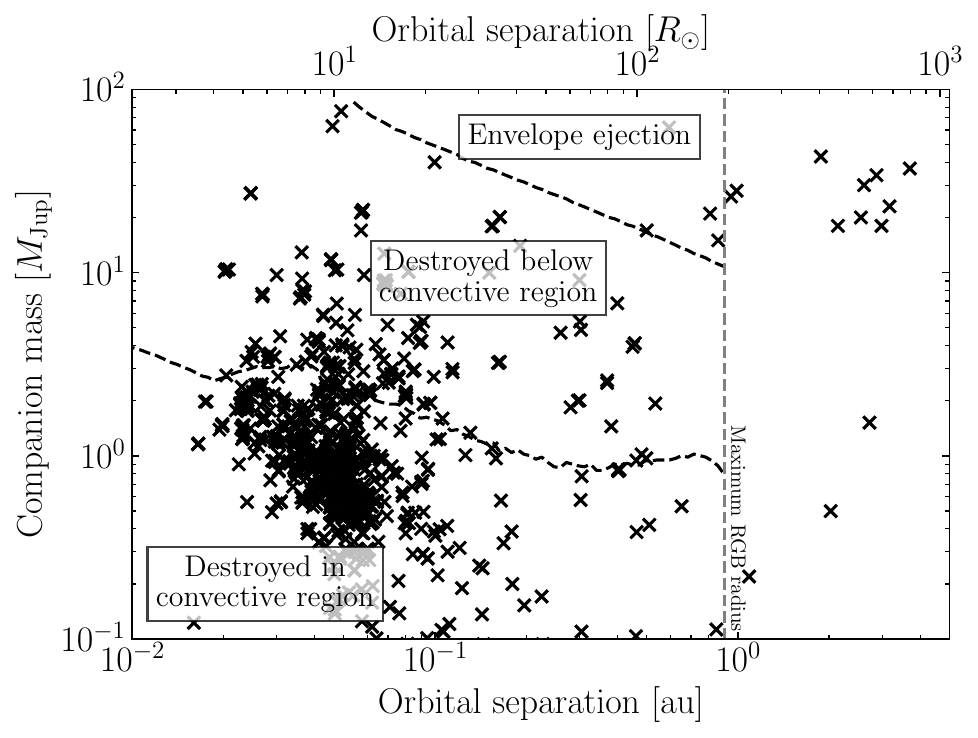}
    \caption{Engulfment outcomes for known substellar bodies \citep[SBs,][]{exoplanetarchive}, as a function of SB mass and orbital separation. This figure assumes that SBs are engulfed at their current separations and that they all orbit \( 1M_\odot \) stars. At a fixed SB mass, envelope ejection becomes easier as the star ascends the red giant branch because its binding energy decreases. A large fraction of SBs engulfed early in post-main-sequence evolution will dissolve in the convective region, yielding potential surface abundance enhancements of the \( ^7 \)Li isotope. We also show systems with orbital separations larger than the maximum stellar radius during the RGB; if tidal forces lead to the engulfment of massive brown dwarfs with orbital separations larger than this maximum RGB radius (i.e., the brown dwarfs in the upper right region of the figure), they could eject the stellar envelope.}\label{fig:engulfment_outcomes}
\end{figure}

We compute \( \dot{E} \) from the work per unit time done by the drag forces,
\begin{equation}
    \dot{E} = - \mathbf{F}_\text{drag}\cdot \mathbf{v}.
    \label{eq:edot}
\end{equation}
We compute the energy transport timescale by adding the cell crossing times in the stellar profile. The most effective energy transport mechanism at each cell (either radiative diffusion or convection) determines the crossing time for that cell.

Figure~\ref{fig:luminosity} shows the quantities involved in our calculation of the average luminosity for the engulfment of SBs of varying masses by a \( 1M_\odot \) star evolved to \( 10R_\odot \). The top left panel compares the orbital decay and energy transport timescales as a function of the orbital separation (equivalently, time). When the orbital decay timescale is much longer than the energy transport timescale, the luminosity is close to the energy deposition rate, whereas deep in the envelope the energy transport timescale is much longer than the orbital decay timescale, making the luminosity much smaller than the energy deposition rate. The bottom left panel shows this behavior. The right panel shows the luminosity.

From \( R_\star=10R_\odot \) to the tip of the RGB, the additional luminosity from engulfing SBs of most masses \( \gtrsim M_\text{Jup} \) is comparable or larger than the stellar luminosity, in some cases by several orders of magnitude (see \citealt{Metzger2017}). SBs engulfed early in the post-main-sequence have higher energy deposition rates (because of the higher stellar density), and the stars that engulfed them are dimmer. On the other hand, evolved stars are sparser and more luminous, but have smaller energy transport timescales at the point of SB destruction. While these luminosity estimates rely on Equation~\ref{eq:lum} (which doesn't accurately describe radiative transfer inside the star) and depend on uncertain processes (such as the destruction of the SB), they suggest that engulfment is energetically significant throughout the post-main-sequence.

After SB destruction, the timescale for the luminosity to return to its original value is roughly the energy transport timescale at the point of destruction. The energy transport timescale at the point of destruction becomes shorter as the star ascends the RGB\@. For a given star, more massive SBs, which tend to also be denser as a result of the mass-radius relation, will result in a longer increase in luminosity because they survive deeper into the envelope. For a Sun-like star evolved to \( 10R_\odot \), the time it takes for the star to return to its original luminosity ranges from \( {\sim}\qty{1}{\year} \) for a \( 1M_\text{Jup} \) planet to \( {\sim}\qty{5000}{\year} \) for a \( 80M_\text{Jup} \) brown dwarf, as shown in Figure~\ref{fig:luminosity}. On the other hand, for a model of the Sun at the tip of the RGB, the time ranges from \( {\sim}\qty{1}{\year} \) to \( {\sim}\qty{25}{\year} \) for the same range of SB masses.

\subsection{Engulfment outcomes}
Figure~\ref{fig:engulfment_outcomes} shows known SBs \citep{exoplanetarchive} as a function of their mass and present-day orbital separation. The dashed line near the top-right corner shows the minimum SB mass for envelope ejection, assuming all deposited energy contributes to the ejection. Since the trajectories of some of the massive SBs that can eject the envelope are likely eccentric (Figure \ref{fig:compare_trajectories_brown_dwarf}), we compute this line using the analytical values for the drag coefficients. For destroyed SBs, the figure shows whether they'll be destroyed in the convective zone or below it, according to the analytical destruction estimates of Equations \ref{eq:disruption:tides} and \ref{eq:disruption:ram}. This figure assumes SBs are engulfed at their present-day orbital separations, and that all SBs orbit \( 1M_\odot \) stars (the average stellar mass reported for these SBs' planetary systems is \( 1.12M_\odot \), with a standard deviation of \( \sim17\% \)).

Under these assumptions, Figure~\ref{fig:engulfment_outcomes} suggests massive SBs can eject the envelopes of evolved Sun-like stars through the transfer of orbital energy. This figure also shows that a substantial fraction of known SBs might be destroyed in the convective region of their host stars, particularly those at closer orbital separations because they are engulfed when the star is more compact and disrupts them more easily. The \( ^7 \)Li contained in these SBs could be carried via convection to the surface, resulting in enhanced surface abundances. However, the mean molecular weight of the SB's enriched material is much higher than that of its surroundings. Some of that material could settle in a layer near the base of the convective region and eventually reach the radiative core \citep{Vauclair2004,Jia2018}. Intermediate-mass SBs (those in the central region of the plot) are massive enough to survive below the base of the convective region, perhaps resulting in opacity changes detectable through asteroseismology.

\section{Conclusions}\label{sec:conclusion}
We studied the engulfment of substellar bodies (SBs) by evolved stars using hydrodynamical simulations of the flow in the vicinity of an engulfed SB (the ``wind tunnel'' framework, schematically depicted in Figure~\ref{fig:setup}). The steps in our numerical framework are:
\begin{enumerate}
\item Determine the hydrodynamical parameter space for planetary engulfment. In particular, the range of values for the dimensionless parameters that affect the morphology of the flow around an engulfed SB.
\item Run hydrodynamical simulations that span this parameter space, characterize the resulting morphologies, and measure the drag coefficients for the drag forces acting on the SB\@.
\item Use the drag coefficients to integrate the equation of motion of an engulfed SB, and estimate observational signatures and outcomes of engulfment.
\end{enumerate}

Some of our main findings are:
\begin{itemize}
\item The interactions of engulfed SBs with their environment are geometrical (ram pressure drag) and gravitational (gravitational drag). Geometrical interactions become increasingly important throughout engulfment (Figure~\ref{fig:flow_params}, Figure~\ref{fig:parameter_space}). All SBs are in the geometrical regime when they are destroyed or when they eject the envelope (Figure~\ref{fig:eject}).
\item According to semi-analytical estimates, tidal disruption is the dominant destruction process for SBs with masses \( \gtrsim 1M_\text{Jup} \) at most stages of stellar evolution.
\item Many SBs will be in the gravitational regime at the onset of engulfment, and transition to the geometrical regime as their orbit shrinks. Our hydrodynamical simulations characterize this transition regime qualitatively (Figure~\ref{fig:wind_tunnel_grid}) and quantitatively (Figure~\ref{fig:drag_coefficients}).
\item The engulfment of an SB could increase the luminosity of a star by up to several orders of magnitude. The time is takes for the star to return to its original luminosity depends on its evolutionary stage and the mass of the SB (Section~\ref{sec:luminosity}; Figure~\ref{fig:luminosity}).
\item Massive SBs could eject the stellar envelope via transfer of orbital energy, if the transferred energy can be efficiently distributed within the envelope (Figure \ref{fig:engulfment_outcomes}). Small SBs are destroyed in the convective region.
\end{itemize}

We discussed the applicability of the wind tunnel framework to planetary engulfment in Section~\ref{sec:cewt}. As implemented in this work, the framework assumes that the density gradient in the stellar envelope and the velocity of the SB are perpendicular, and that the SB moves at approximately the circular Keplerian speed. These assumptions do not hold for some massive SBs, whose orbits develop significant eccentricities during engulfment. For this reason, we used analytical drag coefficients when computing the minimum mass required for envelope ejection. The evolution of the internal structure of the SB and the star remains a significant uncertainty in our calculations. We model the SB as a rigid body with a reflective boundary; while we approximately account for tidal and ram pressure disruption to determine the location in the star where the SB will be destroyed, only hydrodynamical models can describe these processes in detail. Future work could study the evolution of the SB's internal structure to determine the conditions and timescales associated with its destruction. Simulations of the entire star can help understand its response to engulfment. Here we used a simplified model for energy transport to estimate the engulfment luminosity; more sophisticated stellar models including radiative transfer could constrain these observational signatures. The numerical framework we introduced here can be used to study the dynamics of planetary engulfment using comparatively inexpensive simulations that capture the physics of the flow at the scale of the SB\@.

\begin{acknowledgments}
    We thank the anonymous referee for feedback that improved many aspects of this work. We thank Brant Robertson, Ruth Murray-Clay, and Yufeng Du for discussions, and Arcelia Hermosillo-Ruiz for discussions about REBOUND\@. We acknowledge use of the lux supercomputer at UC Santa Cruz, funded by NSF MRI grant AST 1828315, and thank Brant Robertson and Josh Sonstroem for help using lux. The FLASH code was in part developed by the DOE NNSA-ASC OASCR Flash Center at the University of Chicago. R.Y. is grateful for support from a Doctoral Fellowship from the University of California institute for Mexico and the United States (UCMEXUS) and the Consejo Nacional de Ciencia y Tecnolog\'{\i}a (CONACyT), a Texas Advanced Computing Center (TACC) Frontera Computational Science Fellowship, and a National Aeronautics and Space Administration (NASA) Future Investigators in NASA Earth and Space Science and Technology (FINESST) Fellowship (award 21-ASTRO21-0068). R.W.E. is supported by the National Science Foundation Graduate Research Fellowship Program (Award \#1339067), the Heising--Simons Foundation, and the Vera Rubin Presidential Chair for Diversity at UCSC\@. A.~M-B is supported by NASA through the NASA Hubble Fellowship grant HST-HF2-51487.001-A awarded by the Space Telescope Science Institute, which is operated by the Association of Universities for Research in Astronomy, Inc., for NASA, under contract NAS5-26555. E. R-R thanks the Heising--Simons Foundation and the NSF (AST-1911206, AST-1852393, and AST-1615881) for support. M.M. acknowledges support from the National Science Foundation under Grant No. 1909203. Any opinions, findings, and conclusions or recommendations expressed in this material are those of the authors and do not necessarily reflect the views of the NSF\@.
\end{acknowledgments}

\software{FLASH 4.6.2 \citep{Fryxell2000,Dubey2014,Dubey2015}, GNU Scientific Library 2.7 \citep{gsl}, HDF5 \citep{hdf5}, matplotlib 3.6.2 \citep{matplotlib}, MESA \citep{Buchler1976,Fuller1985,Iglesias1993,Oda1994,Saumon1995,Iglesias1996,Itoh1996,Langanke2000,Timmes2000,Rogers2002,Irwin2004,Ferguson2005,Cassisi2007,Chugunov2007,Cyburt2010,Potekhin2010,Paxton2011,Paxton2013,Paxton2015,Paxton2018,Paxton2019}, MESA SDK \citep{Townsend2021}, numpy 1.23.5 \citep{Harris2020}, py\_mesa\_reader \citep{pymesareader}, scipy 1.9.3 \citep{scipy}, unyt 2.9.3 \citep{Goldbaum2018}, yt 4.1.2 \citep{Turk2011}.}

\appendix
\section{Numerical tests}\label{sec:appendix:numerics}
\subsection{Wind tunnel}\label{sec:appendix:numerics:windtunnel}
\begin{figure}
\includegraphics[width=\columnwidth]{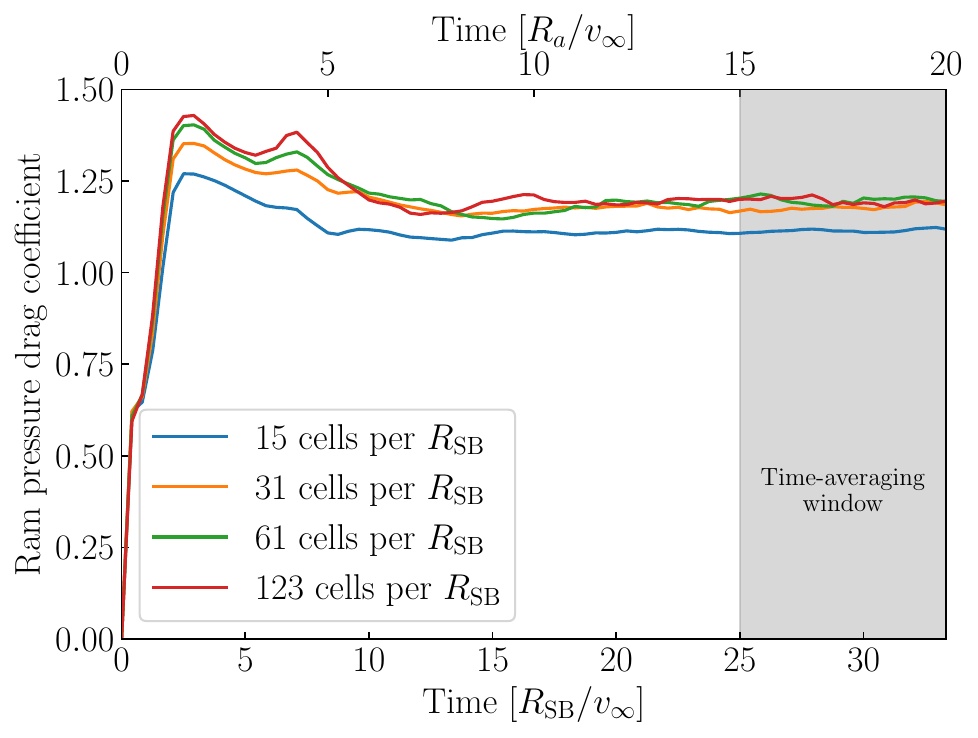}
\caption{Ram pressure drag coefficient as a function of time, for a simulation with \( R_\text{SB}/R_a=0.6 \) and \( \varepsilon_\rho=1 \). Each line corresponds to a different resolution at the surface of the object. We compute the drag coefficient for a simulation by time-averaging the time-dependent drag coefficient between \( 15\max\lp R_\text{SB},R_a\rp/v_\infty \) and \( 20\max\lp R_\text{SB},R_a\rp/v_\infty \).}\label{fig:drag_convergence}
\end{figure}

Figure \ref{fig:drag_convergence} shows the ram pressure drag coefficient in a simulation with \( \varepsilon_\rho=1 \) and \( R_\text{SB}/R_a=0.6 \) at several resolutions. We chose a value of $R_\text{SB}/R_a$ for which computing the ram pressure drag would be the hardest numerically. If $R_\text{SB}>R_a$, the SB is a larger fraction of the domain size (\(L_\text{domain}=10\max\lp R_\text{SB}, R_a\rp\)), so it is easier to resolve. On the other hand, if $R_\text{SB}\ll R_a$, the pressure field around the SB is spherically symmetric (Section \ref{sec:morphology}). The hardest simulations in which to measure ram pressure drag are those with intermediate values of $R_\text{SB}/R_a$, for which $R_\text{SB}$ is small compared to the length of the domain, but for which the pressure around the SB is still asymmetric. The maximum resolution in our test has 122 cells per \( R_\text{SB} \). The ram pressure drag coefficient of a simulation with 31 cells per \( R_\text{SB} \) had a relative error (compared to the simulation with the highest resolution) of \( \approx1.5\% \). We set the refinement in all our simulations such that there are at least 31 cells per \( R_\text{SB} \).

\begin{figure}[t]
\centering
\includegraphics[width=\columnwidth]{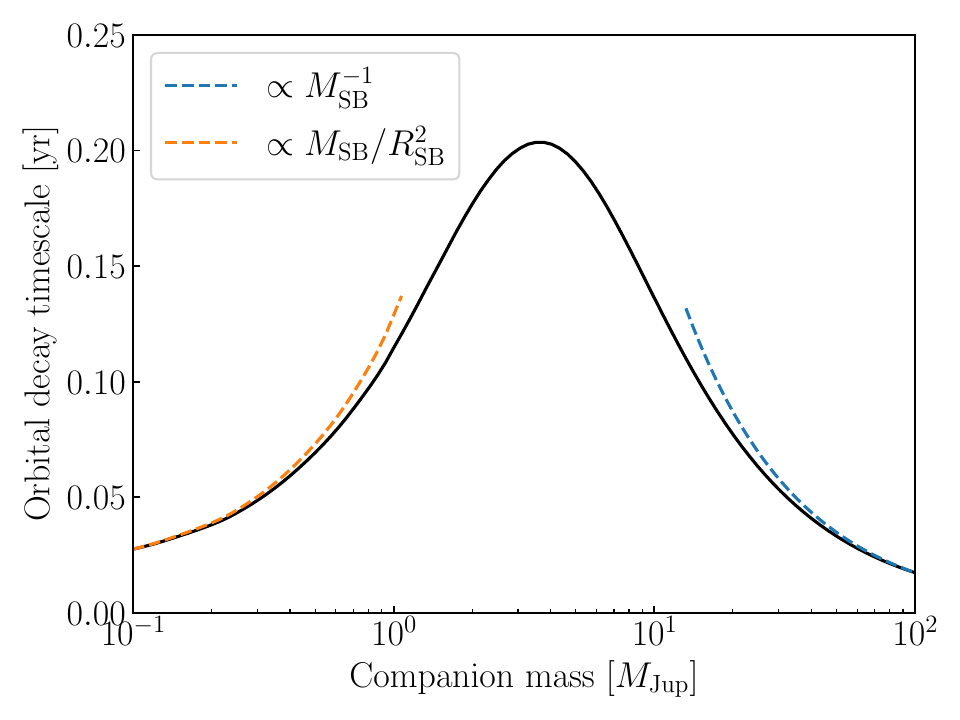}
\caption{Orbital decay time for substellar bodies as a function of their mass, inside a model of a \( 1M_\odot \) star evolved to \( 10R_\odot \).}\label{fig:two_regimes}
\end{figure}

\subsection{Equation of motion integration}
We tested the implementation of the two effects we account for that deviate from the standard two-body problem: the change of the mass enclosed by the orbit of the SB, and the drag forces.

We integrated the orbit of a \( 1M_\text{Jup} \) companion inside a star of radius \( R_\star=10R_\odot \) without drag forces. The initial radial coordinate of the SB is \( 9R_\odot \), and its initial speed is \( 70\% \) of the circular Keplerian speed. After \( 10^4 \) dynamical times of the star, the specific orbital energy of the companion
\begin{equation}
E_\mathrm{orb,sp} = \frac{v^2}{2} - \frac{G M_\text{enc}}{r} + G \int_{M_\text{enc}}^{M_\star}\frac{dM'_\text{enc}}{r}
\end{equation}
was conserved to within a fractional error \( 5\times10^{-7} \).

We then integrated trajectories inside the same star for SBs of different masses, including drag forces (with drag coefficients \( C_g = 1 \) and \( C_p = 0.25 \)). We initially place the SB in a circular Keplerian orbit at \( 9R_\odot \). For each SB mass, we computed the corresponding SB radius using the mass-radius relation, as discussed in Section~\ref{sec:landscape}. Figure~\ref{fig:two_regimes} confirms that, for this star, in the low and high mass limits the orbital decay timescale scales as \( M_\text{SB}/R_\text{SB}^2 \) and \( M_\text{SB}^{-1} \), respectively (see equations~\ref{eq:t_decay_ram} and~\ref{eq:t_decay_grav}). The work done by drag forces was equal to the change in orbital energy to within a fractional error \( 2\times10^{-6} \).

\section{Fitting formul\AE{}}
We fit the results of Figure \ref{fig:engulfment_outcomes}. The minimum companion mass required to eject the envelope of a \( 1M_\odot \) star as a function of its radius is
\begin{equation}
    \frac{M_\text{SB}}{M_\text{Jup}}=\frac{1.3589 x^2+2.1182 x-196.53}{5.0709\times10^{-4} x^3+0.032124 x^2-0.51638 x+1},
\end{equation}
where $x\equiv R_\star/R_\odot$. This formula is valid when $11.45\leq R_\star/R_\odot\leq 193.7$ and agrees with the Figure to within $1.1\%$.

The minimum mass for a companion to survive below the base of the convective zone is
\begin{equation}
    \frac{M_\text{SB}}{M_\text{Jup}}=\lp b_3 x^3 + b_2 x^2 + b_1 x + 1 \rp^{-1}\sum_{i=0}^{5}a_i x^i,
\end{equation}
where $x\equiv\log_{10}\lp R_\star/R_\odot\rp$, $b_1=-3.01905$, $b_2=2.89041$, $b_3=-0.767818$, and
\begin{equation}
\begin{split}
a_i=&\lb5.7773,-23.548,36.3736,\right.\\
&\left.-26.1053,8.96339,-1.20016\rb.
\end{split}
\end{equation}
This formula is valid when $2\leq R_\star/R_\odot\leq 193.7$ and agrees with the Figure to within $7.9\%$.

\section{Data and software availability}\label{sec:appendix:reproducibility}
The software and data required to reproduce our results are available under the digital object identifiers \dataset[10.5281/zenodo.6368227]{https://doi.org/10.5281/zenodo.6368227} and \dataset[10.5281/zenodo.6371752]{https://doi.org/10.5281/zenodo.6371752}, respectively. These repositories include the wind tunnel FLASH setup, the code we used to integrate the equation of motion of the engulfed companion, and the drag coefficients we measured in our hydrodynamical simulations.

\bibliography{bib}

\end{document}